\documentclass{aa}  
\usepackage{graphicx}
\usepackage{txfonts}
\usepackage{hyperref}
\usepackage{booktabs,caption}
\usepackage[flushleft]{threeparttable}
\usepackage[utf8]{inputenc} 
\usepackage[T1]{fontenc}
\DeclareUnicodeCharacter{014F}{\u{o}}
\usepackage{color}

\begin{document} 


   \title{RESIK and RHESSI observations of the 20~September~2002 flare}
\titlerunning{RESIK and RHESSI observations of the 20~September~2002 flare}
   \author{A. Kepa
          \inst{1}
          \and
          R. Falewicz\inst{2}
          \and
          M. Siarkowski\inst{1}
          \and
          M. Pietras\inst{2}
          }

   \institute{Space Research Centre (CBK PAN), Warsaw, Bartycka 18A, Poland\\
              \email{ak@cbk.pan.wroc.pl}
         \and
             Astronomical Institute, University of Wrocław, Kopernika 11, 51-622 Wrocław, Poland\\
             \email{falewicz@astro.uni.wroc.pl}
                         }

   \date{Received 18 May 2020; accepted ...}

 
  \abstract
 {Soft X-ray spectra (3.33~\AA~--~6.15~\AA) from the RESIK instrument on CORONAS-F constitute a unique database for the study of the physical conditions of solar flare plasmas, enabling the calculation of differential emission measures. The two RESIK channels for the shortest wavelengths overlap with the lower  end of the Ramaty High Energy Solar Spectroscopic Imager (RHESSI) spectral energy range, which is located around  3~keV, making it possible to compare both data sets.} {We aim to compare observations from RESIK and RHESSI spectrometers and cross-correlate these instruments. Observations are compared with synthetic spectra calculated based on the results  of one-dimensional hydrodynamical (1D-HD) modelling. The analysis was performed for the flare on 20 September 2002 (\textsf{SOL2002-09-20T09:28}).}{We estimated the geometry of the flaring loop, necessary for 1D-HD modelling, based on images from RHESSI and the Extreme-Ultraviolet Imaging Telescope aboard the Solar and Heliospheric Observatory (SOHO/EIT). The distribution of non-thermal electrons (NTEs) was determined from RHESSI spectra.  The 1D-HD model assumes that non-thermal electrons  with a power-law spectrum were injected at the apex of the  flaring loop. The NTEs then heat and evaporate the chromosphere, filling the loop with hot and dense plasma radiating in soft X-rays. The total energy of electrons was constrained by comparing observed and calculated fluxes from Geostationary Operational Environmental Satellite (GOES) 1~--~8~\AA~data. We determined the temperature and density at every point of the flaring loop throughout the evolution of the flare, calculating the resulting X-ray spectra.}{The synthetic spectra calculated based on the results of hydrodynamic modelling for the 20 September 2002 flare are consistent within a factor~of two with the observed RESIK spectra during most of the duration  of the flare. This discrepancy factor is probably related to the uncertainty  on the cross-calibration between RESIK and RHESSI instruments.}{}

   {}


   \maketitle
%
\section{Introduction}
The majority of published flare models assume that the energy transfer from a magnetic energy release site is by beams of non-thermal electrons (NTEs), which are guided by field lines towards the chromosphere. The chromosphere is then heated and flows back into the flare loop (`chromospheric evaporation'), emitting radiation over a wide range of the electro-magnetic spectrum. The primary energy release sites are commonly taken to be located over the tops of flaring loops. During the precipitation along the magnetic loops, the NTEs are slowed down and thermalized by Coulomb collisions with ambient ions, mostly in the relatively dense plasma near the loop footpoints, but a small part ($\sim$~10$^{-5}$) of their kinetic energy is radiated as hard X-ray (HXR) bremsstralung. The energy delivered by the NTEs also powers heating and macroscopic motions of the upward-moving, evaporated plasma and emission in soft X-rays (SXRs) and chromospheric lines \citep{Brown_1971SoPh...18..489B,Antonucci_1984ApJ...287..917A, Antonucci_1999mfs..conf..331A,Fisher_1985ApJ...289..414F,Fletcher_2011SSRv..159...19F,Holman_2011SSRv..159..107H}. 
When the dense, deeper layers of the chromosphere are heated, the more efficient radiation in optical and UV ranges is observed. Therefore, the deeper into the chromosphere the beam energy is deposited, the less of it is available to drive mass motions into the upper portions of the loop \citep{Mariska_1989ApJ...341.1067M}.

The parameters of the NTE beams, such as the energy distribution, total energy, and precipitation depths, vary rapidly on timescales of seconds. The time variations of the HXR emission are also a function of the temporal and spatial variations of the plasma properties along the flaring loop: mostly the density and the temperature of the plasma near the loop footpoints \citep{Battaglia_2012ApJ...752....4B}. In particular, both HXR and SXR emissions are related to a flux of the NTEs; while the HXR emission is directly excited in a bremsstrahlung process by the NTEs, the SXR emission, thermal in origin, is related to the energy deposited by NTEs in the plasma. Such a model, besides its overall elegancy and self-consistency, leads to important considerations concerning the significance of various auxiliary processes of the energy transport, a total energy budget, and time relations between SXR and HXR emissions \citep{Dennis_1988SoPh..118...49D, Dennis_1993SoPh..146..177D, McTiernan_1999AAS...194.5402M, Falewicz_2009A&A...500..901F}.

In this article, we use the temperatures and densities provided by hydrodynamic modelling of the 20~September 2002 flare, which have been described in detail in previous papers \citep{Falewicz_2011ApJ...733...37F, Falewicz_2015ApJ...813...70F,Siarkowski_2009ApJ...705L.143S},  to reconstruct differential emission measure distributions (DEMs). Based on these DEM distributions, we then calculate the synthetic spectra and  compare  these with observations in the soft X-ray range. For observed spectra we use RESIK  level~2 data (\url{www.cbk.pan.wroc.pl/experiments/resik/RESIK\_Level2/index.html}), which were available for this event.  RESIK \citep{Sylwester_2005SoPh..226...45S} was a high-resolution crystal spectrometer operating in the nominal energy range 2.01~keV~--~3.72~keV (3.33 ~\AA~--~6.15~\AA). It  was built at the Solar Physics Division of the Space Research Centre  of the Polish Academy of Sciences with international collaboration and observed the whole Sun from 2001 to 2003 on board the Coronas-F \citep{Kuznetsov_2014ASSL..400....1K} spacecraft. The crystals used as dispersive elements in RESIK were thin wafers (0.5 mm and 1 mm) of silicon and quartz monocrystals bent to convex cylindrical profiles. This geometry allowed the overall size to be as small as possible, and for whole spectral ranges to be observed simultaneously without scanning motions. The crystals chosen have little or no fluorescence, thereby reducing background radiation which was a problem with a number of previous crystal spectrometers. The spectrometer was calibrated at Rutherford Appleton Laboratory and
Mullard Space Science Laboratory, making RESIK spectra suitable for comparison to data from other instruments, such as for example Geostationary Operational Environmental Satellite -- GOES \citep{Donnelly_1977STIN...7813992D, White_2005SoPh..227..231W} and RHESSI \citep{Lin_2002SoPh..210....3L,Hurford_2002SoPh..210...61H,Smith_2002SoPh..210...33S}. 
 \begin{figure*}[th!]
   \centering
\begin{minipage}[b]{5.5cm}   
    \includegraphics[width=5.5cm]{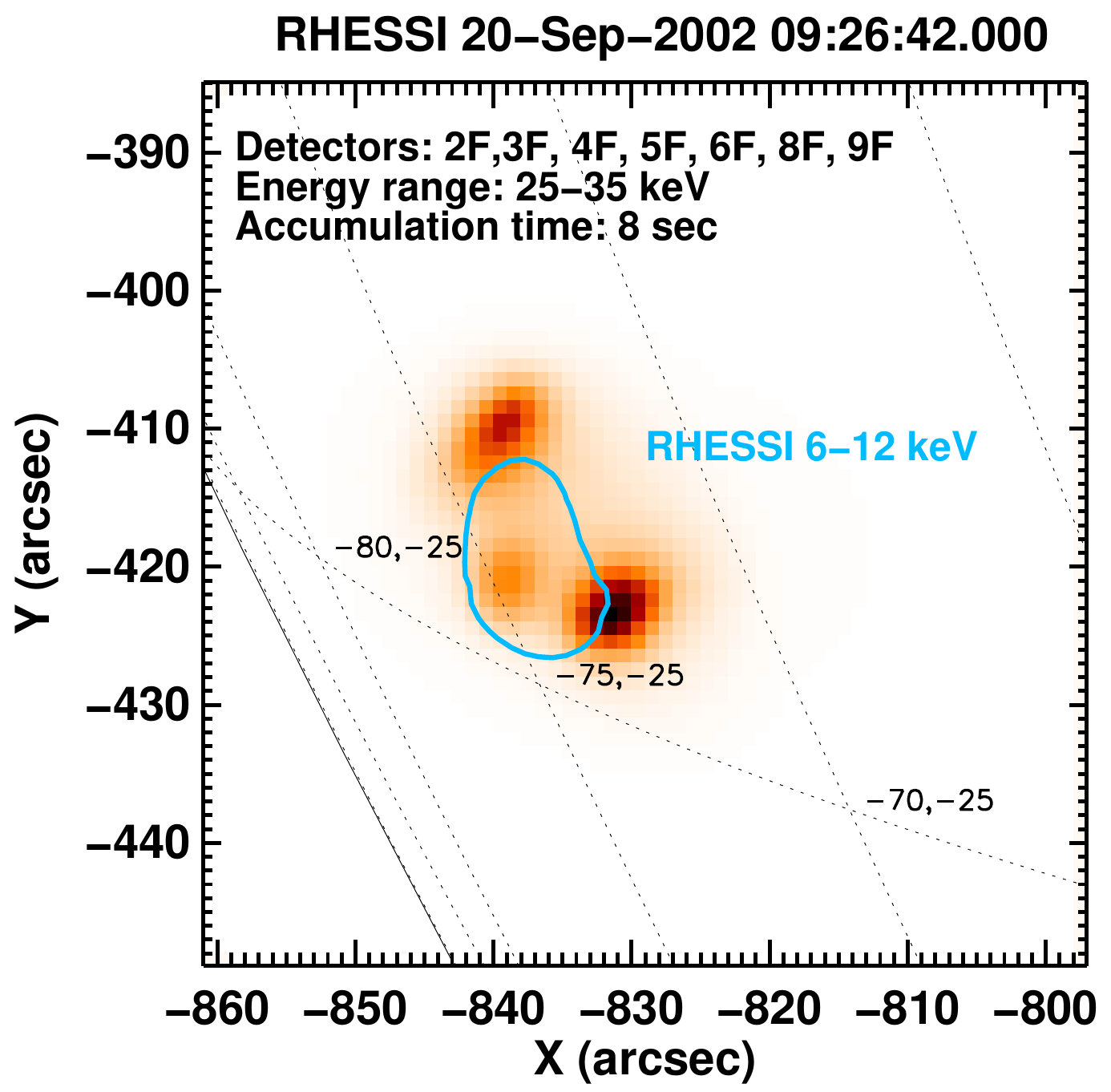}
\end{minipage}
  \centering    
\begin{minipage}[b]{5.5cm}
   \includegraphics[width=5.5cm]{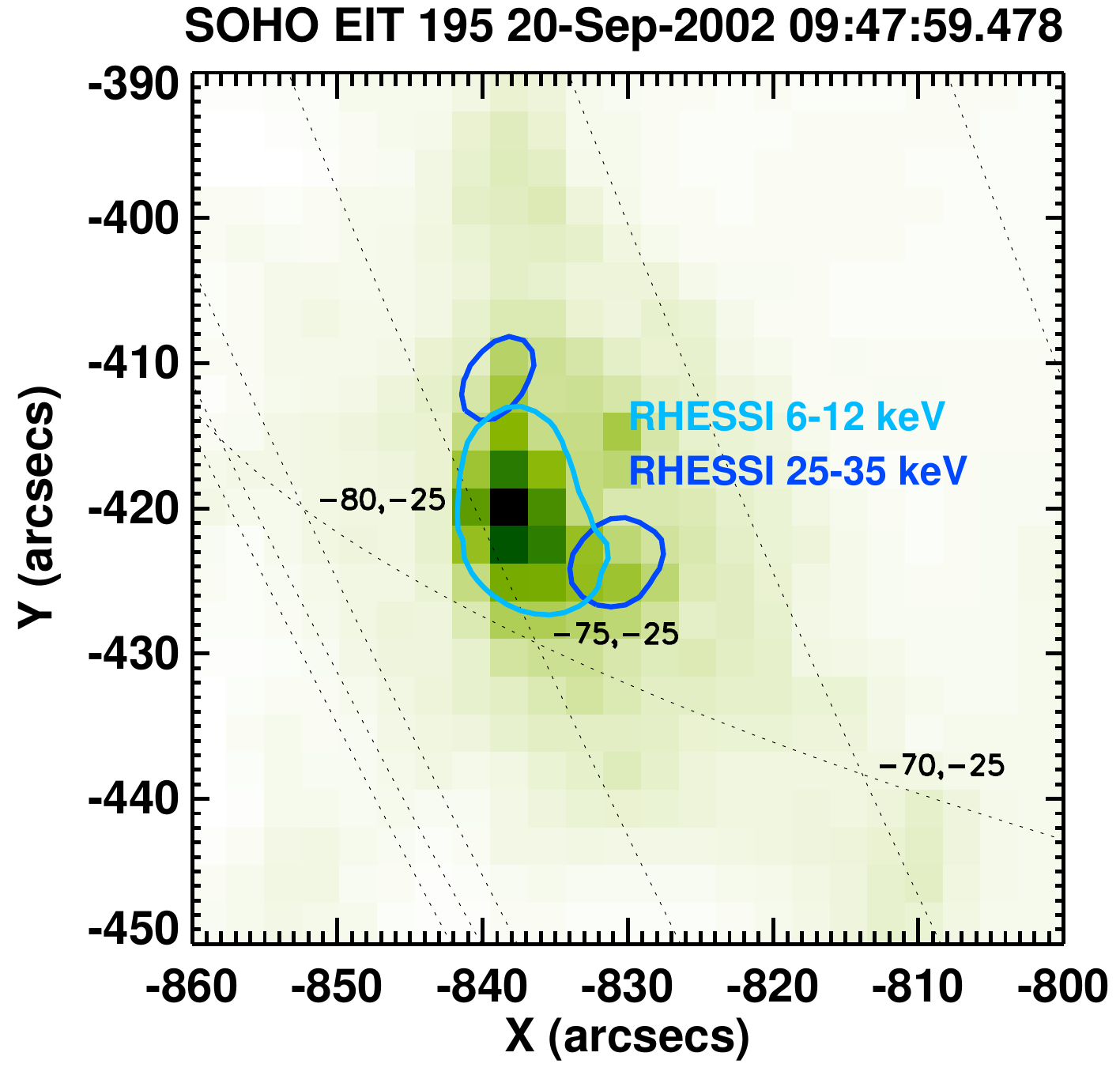}
\end{minipage}
 \centering    
\begin{minipage}[b]{5.5cm}
   \includegraphics[width=5.5cm]{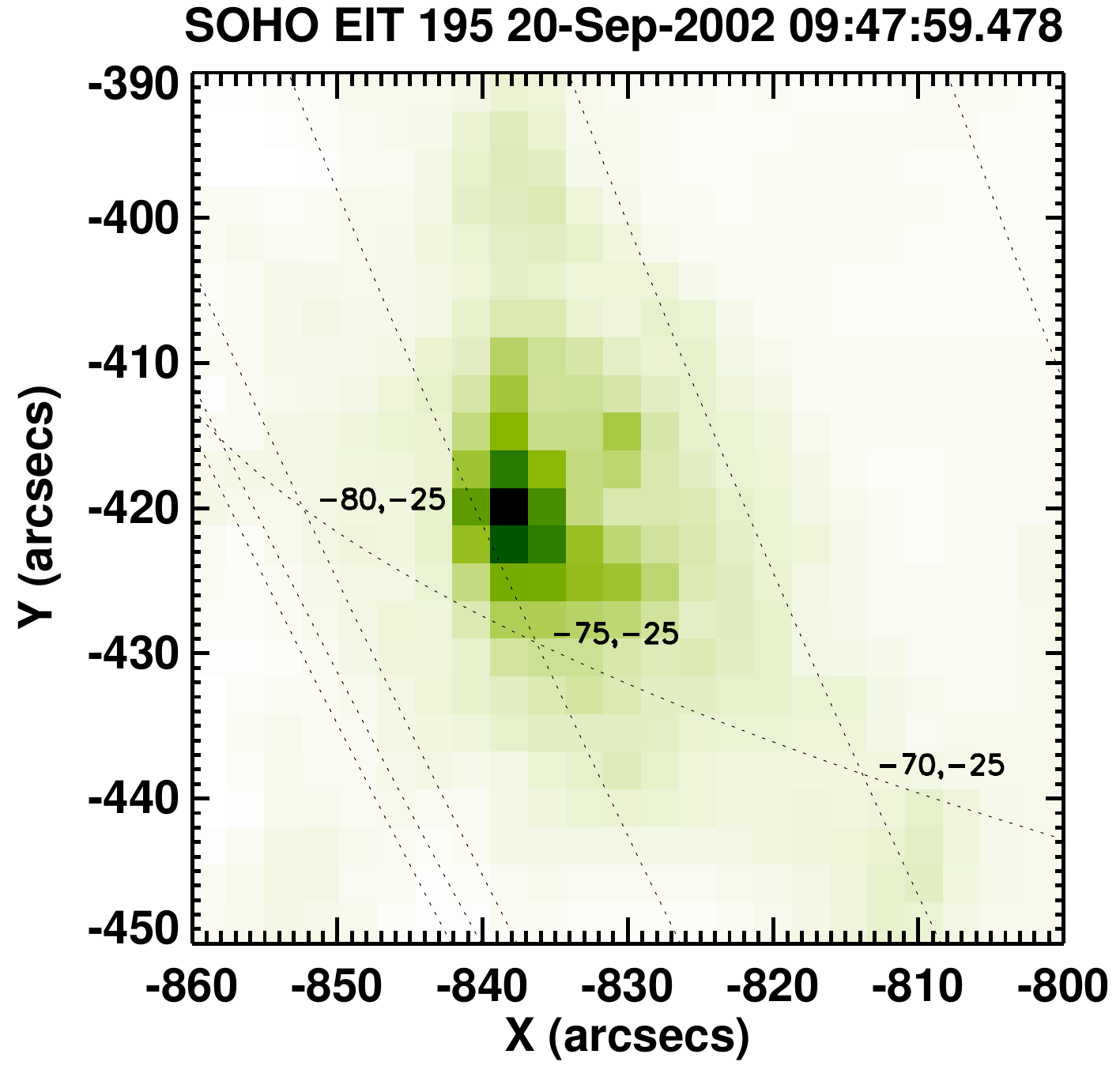}
\end{minipage}
      \caption{RHESSI and EIT images of the M1.8 GOES class solar flare on 20 September 2002. Left panel: RHESSI image restored using the PIXON method in 6~--~12~keV (isophote) and 25~--~35~keV (red scale) energy bands. The signal was accumulated between 09:26:42~UT and 09:26:50~UT \citep{Falewicz_2015ApJ...813...70F}. Middle: SOHO/EIT 195~\AA~image taken at 09:47:59~UT (green  scale) overplotted with the RHESSI 6~keV~--~12~keV and 25~--~35~keV images registered at 09:26:42 UT (dark and light blue contours).The isophotes shown correspond to 30\% of the peak intensity.  Right panel: SOHO/EIT image shown without RHESSI contours.}
         \label{FigVibStab}
   \end{figure*}
The article is organised as follows. A description of instruments and observations is given in Section 2. In Section 3 we discuss numerical models of the flare. Section 4 contains calculations of the differential emission measure distributions and comparison between the observed and calculated spectra. In Section 5 we show results of cross-calibration between RESIK and the Ramaty High Energy Solar Spectroscopic Imager (RHESSI) instruments. A  discussion and conclusions are given in Section 6.
\section{Observations}

RESIK  was a Bragg crystal spectrometer designed to observe solar active region and flare plasmas in SXRs. The orbit of the Coronas-F spacecraft was nearly circular (about 500~km altitude)  and crossed the auroral regions of van Allen belts at least four times with each revolution, as the orbital plane was inclined at an angle of 82.5 degrees to the equator. In addition, Coronas-F penetrated the South Atlantic Anomaly (SAA) approximately six times every 24 hours. When the Coronas-F spacecraft crossed through a polar van Allen radiation belt or  SAA, the high-voltage spectrometer was turned off to prevent damage to electronic devices. Despite these short breaks in observations, RESIK obtained several hundred thousand spectra observed during various phases of the flares, including non-flare intervals. The instrument measured the whole Sun spectra with high spectral resolution in four channels with the following nominal ranges: no. 1: 3.33~\AA~--~3.90~\AA, no. 2: 3.78~\AA~--~4.32~\AA, no. 3:~4.23~\AA~ --~4.92~\AA, and no. 4: 4.90~\AA~--~6.15~\AA. Because of the bent nature of the crystals, all wavelengths were simultaneously recorded. The accumulation time of any given spectrum was a multiple of the basic DGI (data gather interval) which was 2~s. Its length depended on the activity level, changing dynamically with the evolution of an observed flare. 
   \begin{figure}[h!]
   \centering
    \includegraphics[width=8cm]{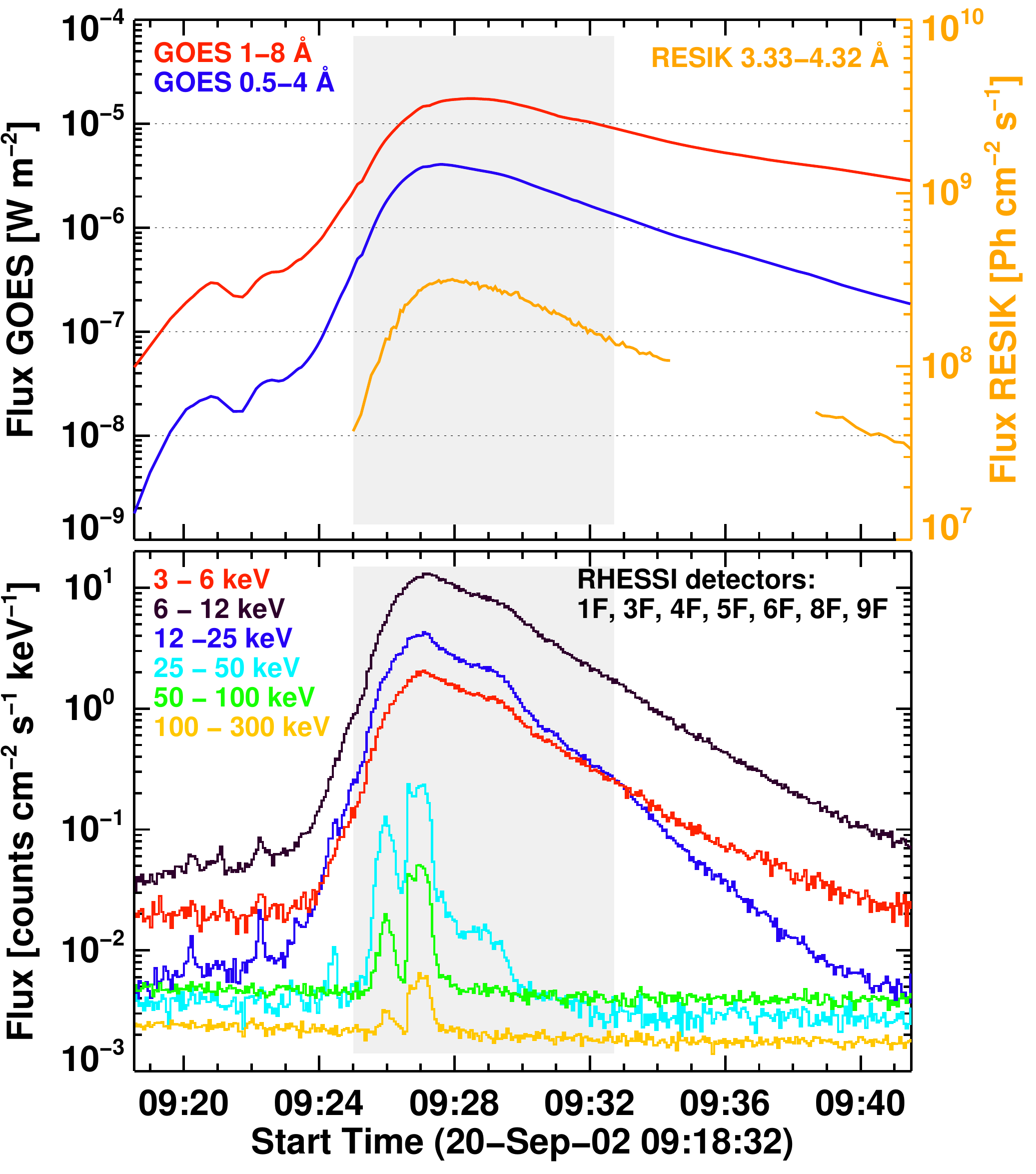}
      \caption{Upper panel:  GOES 0.5 -- 4~\AA~and 1 -- 8~\AA~light curves and temporal evolution of RESIK flux in the 3.33~\AA~--~4.32~\AA~wavelength range. The GOES fluxes should be referred to the left-hand y-axis, the RESIK fluxes to the right-hand y-axis.  Lower panel:  RHESSI light curves of five energy bands between 4 and 300 keV  taken during the 20 September 2002 flare. The grey box indicates the time interval from which the observations were used for the analysis.}
         \label{FigVibStab}
   \end{figure}
In this article we present the results of the analysis for the flare which occurred on 20 September 2002 (maximum phase near 09:28~UT) with GOES class M1.8. RESIK observations are available for two spacecraft  orbits, from 09:25~UT to 09:34~UT and from 09:38~UT to 09:50~UT. The  analysis was performed for the spectra measured in the first time period (about 120 individual observations with time resolutions from 4~s to 14~s), which covered the rise, maximum, and part of the decay phase of the flare. Due to the contribution of emission from higher orders of crystal diffraction and onboard computer setup problems, only spectra from the first two RESIK channels were used for the present study. In the analysed channels (spectral range 3.3 to 4.3~\AA) the emission lines of hydrogen- and helium-like ions such as argon, potassium, and sulphur are observed. All these lines  are formed at high temperatures (above 5~MK).

 This flare was selected because of its simple single loop initial structure apparent in RHESSI images and those from the Extreme Ultraviolet Imaging Telescope onboard Solar and Heliospheric Observatory (SOHO/EIT). This flare was also selected because of the good coverage in time of RESIK, RHESSI, GOES, and SOHO observations (see Figure~2).

The flare on 20 September 2002  was recorded by the RHESSI satellite without activation of the instrument's  attenuators, meaning that the observed spectra do not have any discontinuities. RHESSI observed X-ray emission from the whole Sun over a wide energy range (3~keV~--~17~MeV) with high temporal and energy resolutions as well as with high signal sensitivity, using nine coaxial germanium detectors. Such characteristics allow restoration of the 2D images and spectra in the X-ray band, and provide valuable data for investigation of the thermal and non-thermal emission of the solar flares.

The X-ray emission of the flare was also recorded with the GOES X-ray monitors. 
The GOES X-ray instruments have observed the solar activity for several decades and have been used to produce the largest database of solar flares \citep{Ryan_2012ApJS..202...11R}. On each GOES satellite there are two X-ray sensors (XRSs) that continuously record full-disc integrated X-ray emission in two energy bands, 1~\AA~--~8~\AA~and~0.5~\AA~--~4~\AA,~with 3~s temporal resolution.

The SOHO/EIT telescope provides full-disc images taken in four bands, 171~\AA, 195~\AA, 284~\AA, and 304~\AA,~ with 2.6 arcs  pixel$^{-1}$ spatial resolution. The temperature range of these bands is roughly 8 $\times$ 10$^4$ -- 2 $\times$ 10$^6$ K. For the purpose of modelling, images in the 195~\AA~ filter were used.

The 20 September flare has already been discussed by other authors. \cite{Siarkowski_2009ApJ...705L.143S} investigated the emission of this event in the early beginning of  the impulsive phase of the flare, when the non-thermal electrons are presumed to have been injected into the flaring loop and plasma heating begins. 

\cite{Falewicz_2011ApJ...733...37F} investigated the energy budget of the flare as well as various heating mechanisms for the gradual phase.  These authors showed that for this flare there was no need to use any additional ad hoc heating mechanisms other than heating by non-thermal electrons. The flare was also used as a prototype of a solar flare, which ensures reasonable geometrical and physical parameters of the modelled flaring loop for one-dimensional (1D) hydrodynamic and two-dimensional (2D) magnetohydrodynamic models \citep{Falewicz_2015ApJ...813...70F}.

The 20 September flare occurred in AR NOAA 10126 (S23E69); its RHESSI and GOES light curves are shown in Figure 2. The soft X-ray emission (1~--~8~\AA) of the flare recorded by GOES started to rise at  09:18:15 UT and continued till 10:00 UT. The rise in the higher-energy channel (0.5~--~4\AA) emission started simultaneously but peaked one minute  earlier at 09:27:30 UT. The impulsive phase of the flare recorded by RHESSI in X-rays $\geq$ 25 keV started at 09:25:24 UT and had two maxima around 09:26 UT and 09:27 UT. In the 25 - 50 keV energy range, a small spike of emission was recorded between 09:24:16 UT and 09:24:32 UT. The SXR emission recorded by RHESSI below 25 keV started to rise simultaneously with the SXR emission recorded by GOES \citep{Siarkowski_2009ApJ...705L.143S}. Images of the flare were reconstructed using RHESSI data collected with sub-collimators 2F, 3F, 4F, 5F, 6F, 8F, and 9F, integrated over 8 s periods and the PIXON imaging algorithm with 1 arcsec pixel size \citep{Metcalf_1996ApJ...466..585M, Hurford_2002SoPh..210...61H} was used. This algorithm searches for solutions with the smallest number of degrees of freedom consistent with the observations and does not assume shapes, sizes, or number of emitting sources.  Although compared with other imaging methods, the PIXON method is very time consuming, its main advantage  is that it offers superior noise suppression and accurate photometry. PIXON is the best method for image reconstruction of extended sources in the presence of several compact sources, and its efficacy as has been demonstrated by several authors \citep{Aschwanden_2004SoPh..219..149A, Kontar_2010ApJ...717..250K,Chen_2012ApJ...748...33C}. An accumulation time of 8 seconds was sufficient to maintain good statistics to reconstruct images in the 25~--~35 keV energy range.

The reconstructed images revealed that SXR emission in 6~--~12~keV and intermediate 12~--~25~keV energy bands are coincident with the flare location of the hard emission in 25~--~35~keV. These observations also indicate that SXR emission recorded by GOES during the early phase of the flare  originated from the analysed event. The images obtained in energies above 25 keV show two foot-points and a loop-top source characteristic of a single flaring loop (see Figure 1). Inspection of the images using a method proposed by \cite{Aschwanden_1999ApJ...517..977A} allowed us to determine the main geometrical parameters of the flaring loops, necessary for hydrodynamic modelling. The cross-section of the loop S = (9.0$\pm$7.6)$\times$ $10^{16}$ cm$^2$ was estimated as an area of the structure delimited by a flux level equal to 30\% of the maximum flux in the 25 -- 35 keV energy range. The cross section was assumed to be constant during flare modelling. The half-length of the loop L$_{\textrm{0}}$ = (9.3$\pm$1.1)$\times$ $10^8$ cm was estimated from a distance between the centres of gravity of the foot-points, assuming a semi-circular shape of the loop. Images obtained with the SOHO/EIT telescope in the 195~\AA~band at 09:47:59~UT and 9:59:59~UT (after the impulsive phase of the flare) confirmed the single-loop structure of the flare (see Figure 1, right panel). 

The examples of the  RHESSI spectra  for this event are shown in Figure 3. The average RESIK spectrum observed in the first two channels with wavelength ranges selected for further analysis is shown in Figure 4 in the section describing the calculations of differential emission measure distributions.  

\begin{figure*}[]
 \begin{minipage}[t]{0.5\linewidth}   
    \includegraphics[width=8cm]{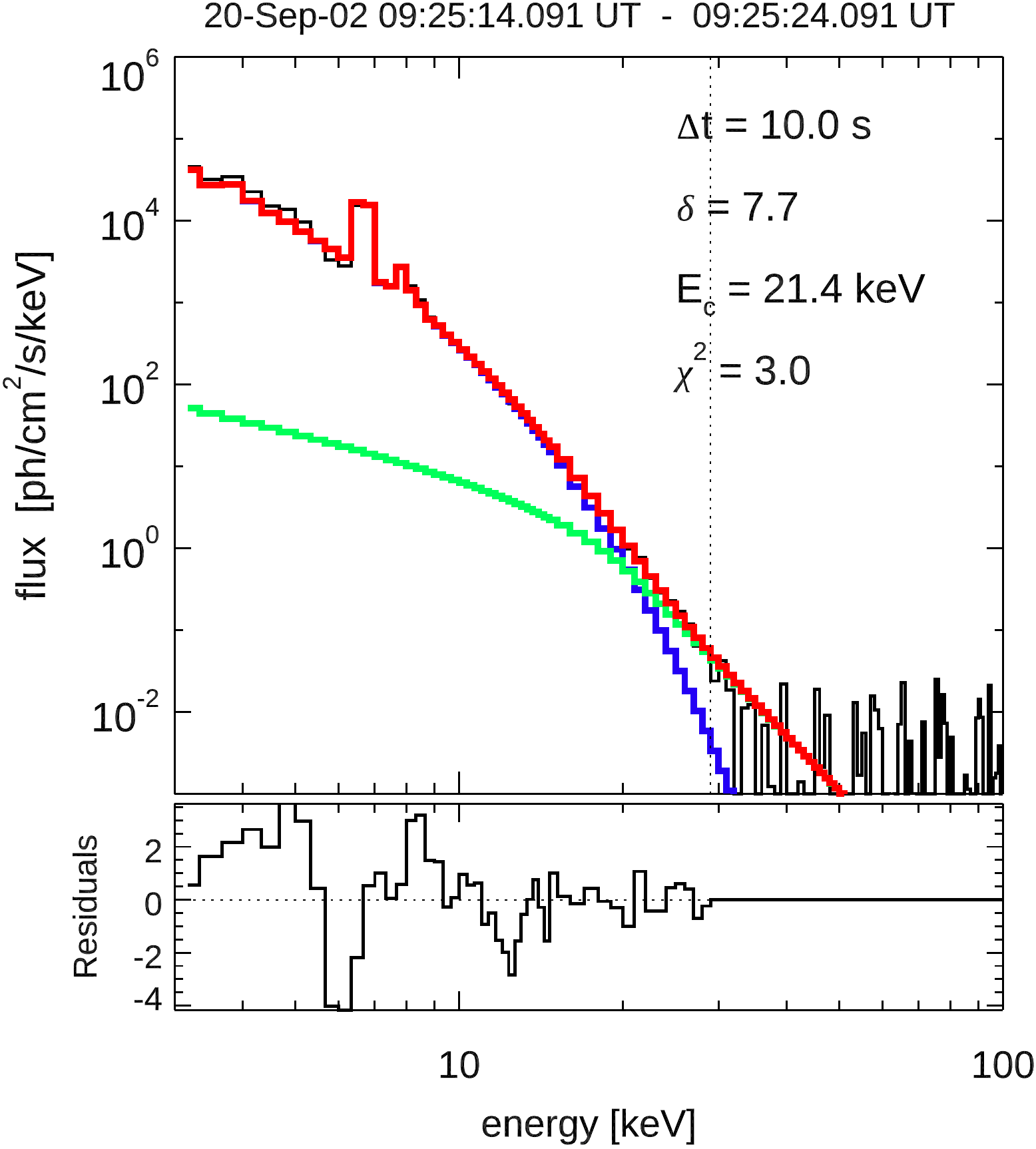}
\end{minipage}   
\begin{minipage}[t]{0.5\linewidth}
   \includegraphics[width=8cm]{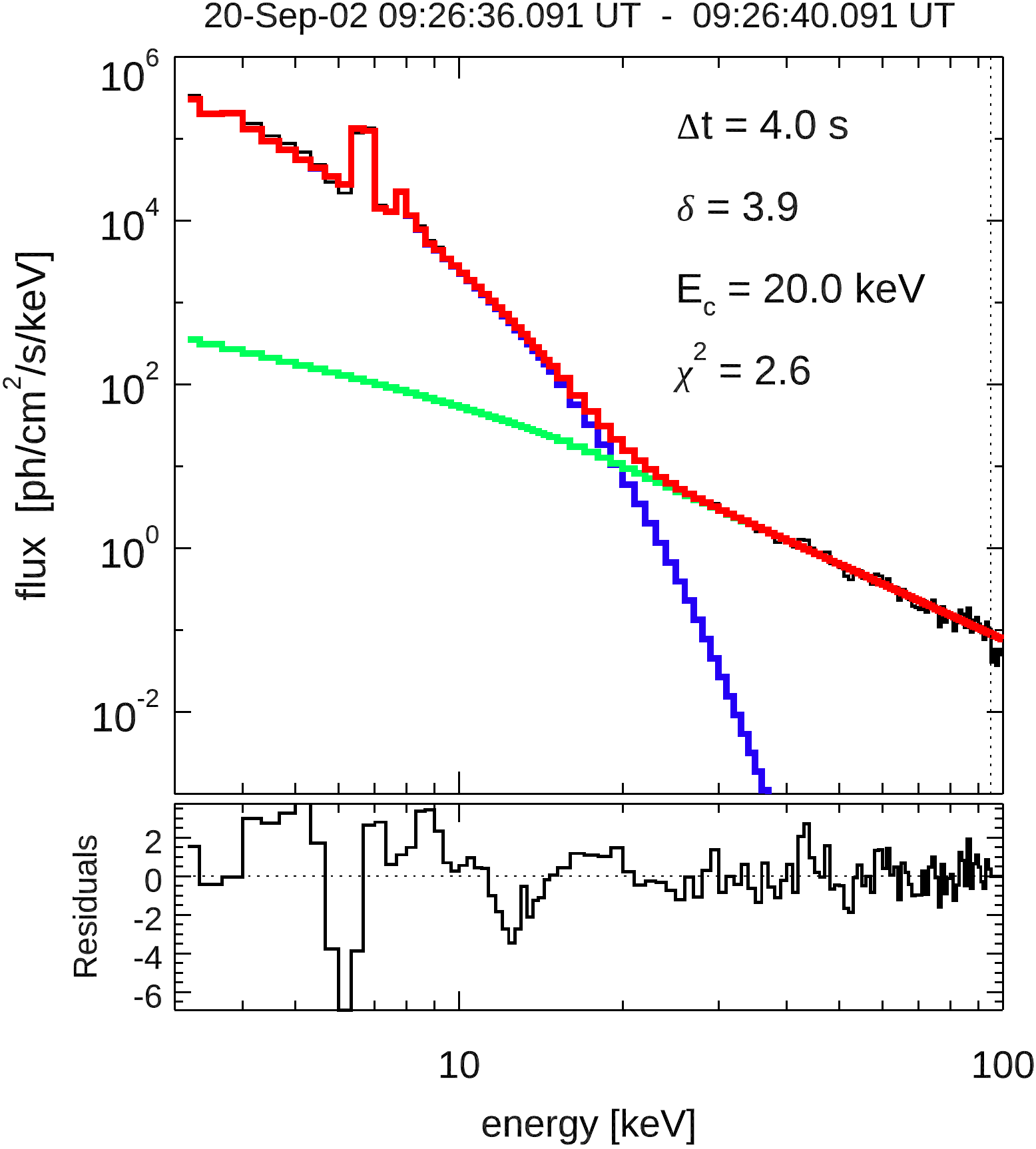}
\end{minipage}
\begin{minipage}[t]{0.5\linewidth}   
    \includegraphics[width=8cm]{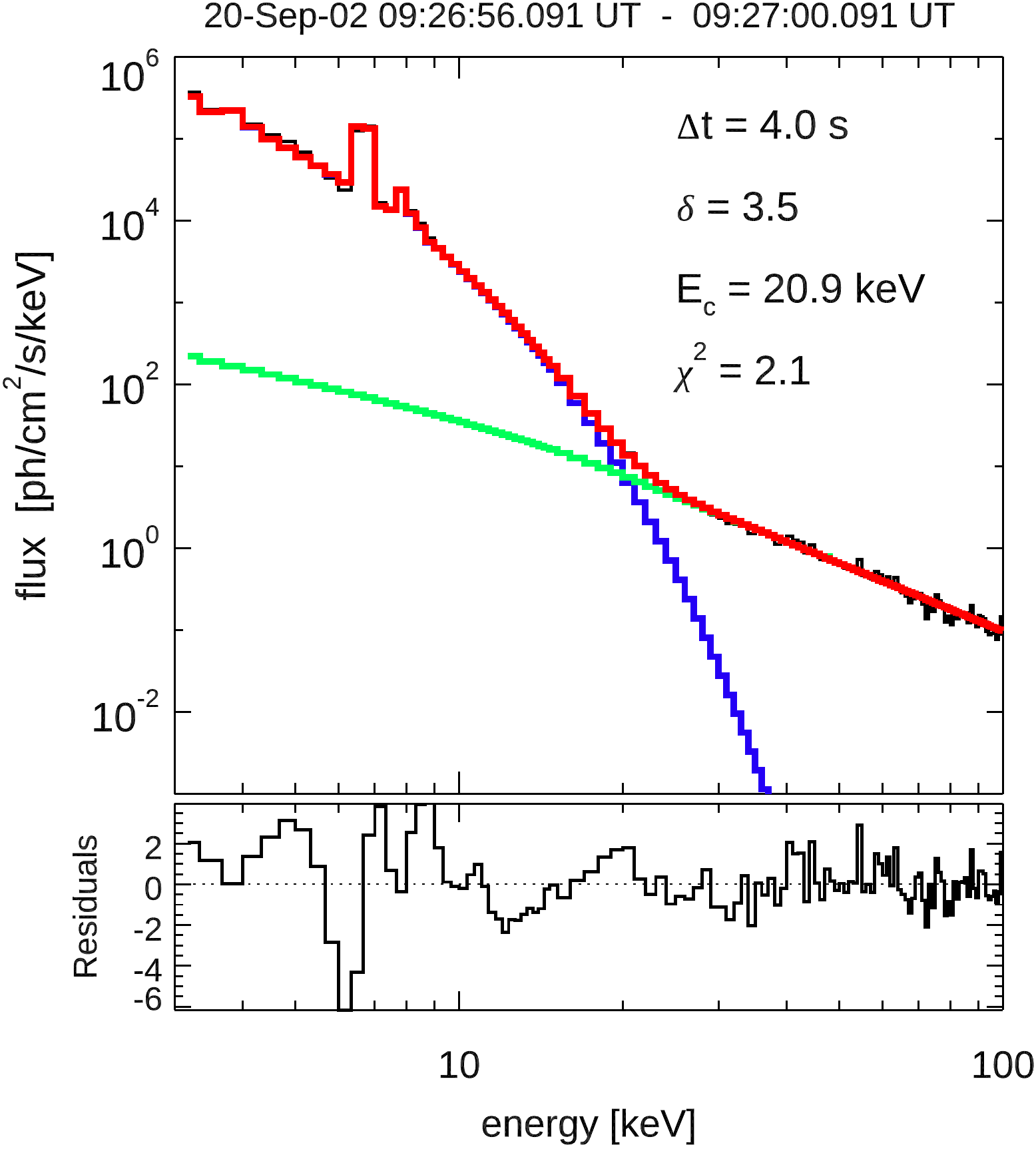}
\end{minipage}
 \begin{minipage}[t]{0.5\linewidth}
H   \includegraphics[width=8cm]{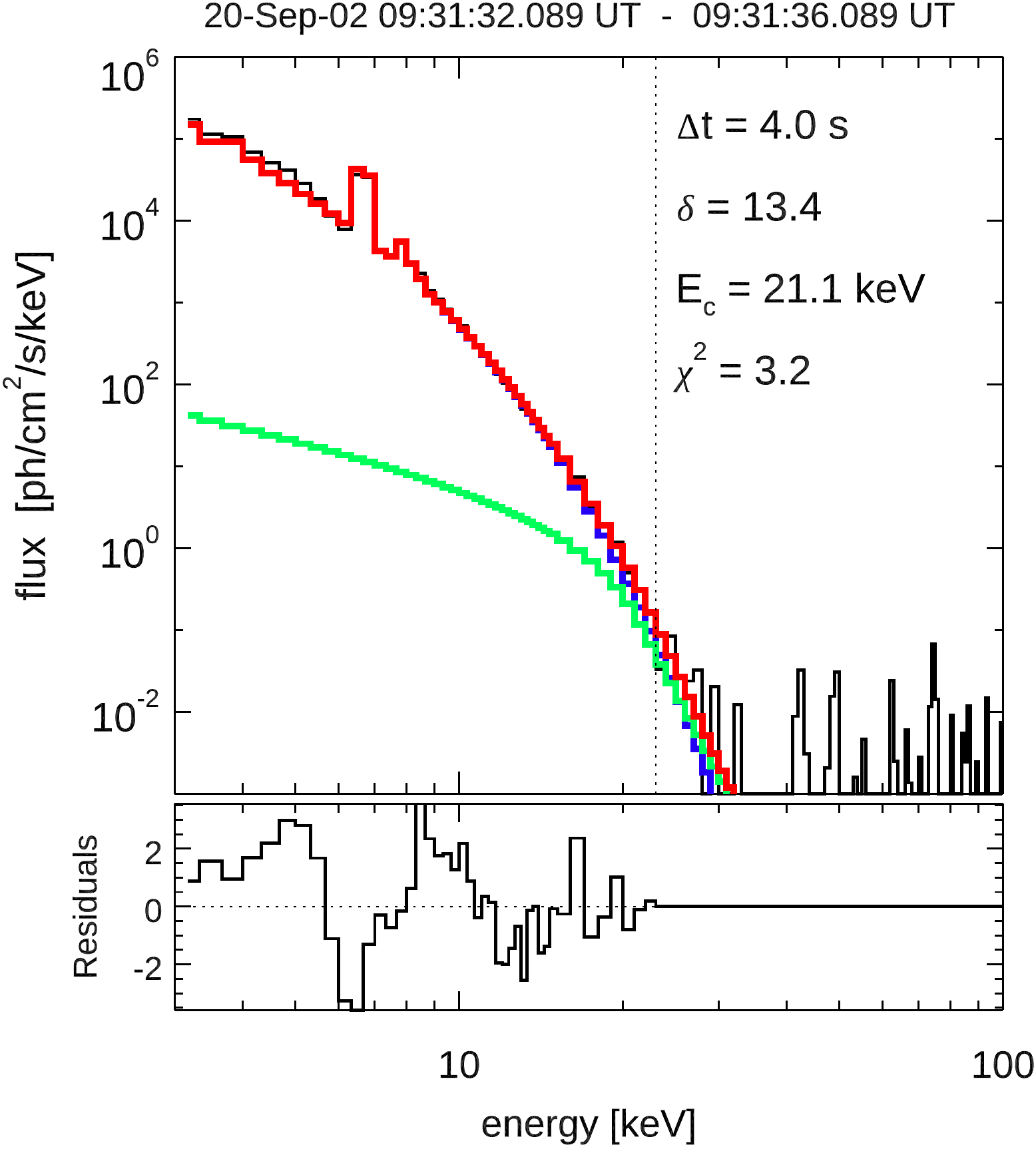}
\end{minipage}
      \caption{RHESSI spectra taken in four time periods of the flare on 20 September 2002. The spectra were fitted with an isothermal model (blue colour) and thick-target model (green). The combined spectra are shown in red. The accumulation time ($\Delta$t),  power law index of the electron energy distribution ($\delta$), low-energy cutoff of the electron distribution (E$_{\textrm{c}}$), and $\chi^2$ values are indicated in the top right-hand corner of each panel.  The vertical dashed lines correspond to the energy values (upper limits) to which the spectra have been fitted and are 29~keV, 95~keV, 165~keV and 23~keV for the spectra shown in the figure. The lower limit is~3~keV for each analysed spectrum.}
         \label{FigVibStab}
   \end{figure*}      
\section{Numerical model of the flare}

The 1D–HD numerical model of the flare on 20 September 2002 was calculated using the modified hydrodynamic 1D Solar Flux Tube Model \citep{Mariska_1982ApJ...255..783M, Mariska_1989ApJ...341.1067M, Falewicz_2009A&A...500..901F}. It was assumed that the flare plasma was heated only by time-variable NTE beams, as has been found relevant for some flares by \cite{Falewicz_2011ApJ...733...37F} and \cite{ Falewicz_2014ApJ...789...71F}. However, a similar problem was considered by other authors, such as for example \cite{Nagai_1984ApJ...279..896N}, \cite{Fisher_1985ApJ...289..434F, Fisher_1985ApJ...289..425F,Fisher_1985ApJ...289..414F}, \cite{Liu_2009ApJ...702.1553L},  
 and \cite{Reep_2013ApJ...778...76R}, revealing that non-thermal electron beams are not always responsible for heating the flare plasma, and that often some additional energy source is needed \citep{Liu_2013ApJ...770..111L,  Aschwanden_2016ApJ...832...27A}. The numerical model of the flaring loop was calculated using numerous specific geometric and thermodynamic characteristic parameters. The loop length and cross section, the initial pressure at the transition region, and plasma temperature were estimated from RHESSI and GOES observations. 

The applied methods of data analysis and numerical modelling of the flares were similar to those presented in detail in \cite{Falewicz_2011ApJ...733...37F, Falewicz_2015ApJ...813...70F, Falewicz_2017ApJ...847...84F}. Our method is novel in that it uses a revised set  of procedures for  evaluation and optimisation of the low-energy cutoff of the electron distribution E$_{\textrm{c}}$ by comparison of the observed and calculated flux at 1 -- 8~\AA~from the GOES instrument. Compared to previous work, the geometrical parameters of the loop are also better adapted, leading to improved agreement between the numerical model and observations. The half-length of the loop has been set as L$_{\textrm{0}}$= 9.5$\times$10$^{8}$~cm, where the cross section of the flaring loop  was used as S = 2.0$\times$10$^{16}$~cm$^2$. The L$_{\textrm{0}}$ and S values lie well within the ranges determined from the images. The gas pressure value (P) at the base of the transition region and for the beginning of flare modelling was assumed to be P~=~34.4~{dyn~cm$^{\textrm{-2}}$ which allowed us to obtain the best agreement with GOES flux in 0.5~--~4~\AA}. 

The HXR data were analysed using the RHESSI Object SPectral Executive (OSPEX) routine in the SolarSoftWare (SSW) package. The X-ray spectra were measured with 4~s temporal resolution in 158 energy bands ranging from 3 to 300 keV and were corrected for pulse pileup and decimation. The spectra were fitted using isothermal plus thick-target models: Bremsstrahlung ver. 2 (vth + thick2). The thermal model was defined by single temperature and emission measure  values of the optically thin thermal plasma. The thick-target model was defined by the total integrated NTE flux F, the power law index of the electron energy distribution $\delta$, and the low-energy cutoff of the electron distribution E$_{\textrm{c}}$. 

The RHESSI spectra were analysed using a forward and backward automatic fitting procedure, starting from the moment when the non-thermal component was strong and clearly visible, taken to be the maximum of the impulsive phase (see Figure 1). The averaged non-flare background spectra were removed before the fitting procedure. The background spectra for energies below 50 keV were accumulated and averaged from pre-flare periods between 09:00~UT and 09:06~UT. We used a linear interpolation of the background between the time intervals before and after the impulsive phase for energies above 50~keV. The fundamental assumption during flare modelling was that only NTE beams derived from RHESSI spectra delivered energy to the flaring loop (via Coulomb collisions with the plasma filling the loop). The deposition of the energy by NTEs was calculated using a rough analytical approximation of the Focker-Planck formalism given by \cite{Fisher_1989ApJ...346.1019F}. The hydrodynamic evolution of the flaring plasma was modelled with the modified Naval Research Laboratory Solar Flux Tube Model code \citep{Mariska_1982ApJ...255..783M, Mariska_1989ApJ...341.1067M, Falewicz_2009A&A...500..901F}. Initial, quasi-stationary pre-flare models of the flaring loops were built using geometrical (semi-length L$_{\textrm{0}}$, cross-section S) and thermodynamic (initial pressure at the base of transition region P$_{\textrm{0}}$, temperature, emission measure, mean electron density, and GOES-class) parameters estimated from RHESSI and GOES data. In the course of the calculations both the semi-lengths and cross-sections of the loops were refined (within the error range only) in order to obtain the best conformity between theoretical and observed GOES flux in the 0.5~--~4~\AA~energy range. We found that acceptable agreement of the calculated and observed fluxes could be obtained using various values of E$_{\textrm{c}}$ in the range from~12~keV to~40~keV.  

\section{Differential emission measure distributions}

One of the important physical parameters which can be determined based on  observed spectra is 
differential emission measure distribution (DEM, $\varphi(T)$), which characterises the amount of material at temperature $T$ and in volume $V,$ and is defined as:
\begin{equation}
DEM=\varphi(T)=N_e^2\frac{dV}{dT}
,\end{equation}
where $N_e$ is electron density of the plasma.

Differential emission measure distribution convolved with the theoretically calculated emission function $f_i(T,\lambda_i,N_e)$ corresponds to each flux ($F_i$) in an appropriate spectral band/line $i$:
\begin{equation}
F_i=A_i\int_{T} f_i(T,\lambda_i,N_e)\varphi(T) dT
,\end{equation}
where $A_i$ represents the abundance of an element contributing to the flux of a particular line or spectral band $i$.

\begin{figure}[h!]
\centering
\includegraphics[width=9cm]{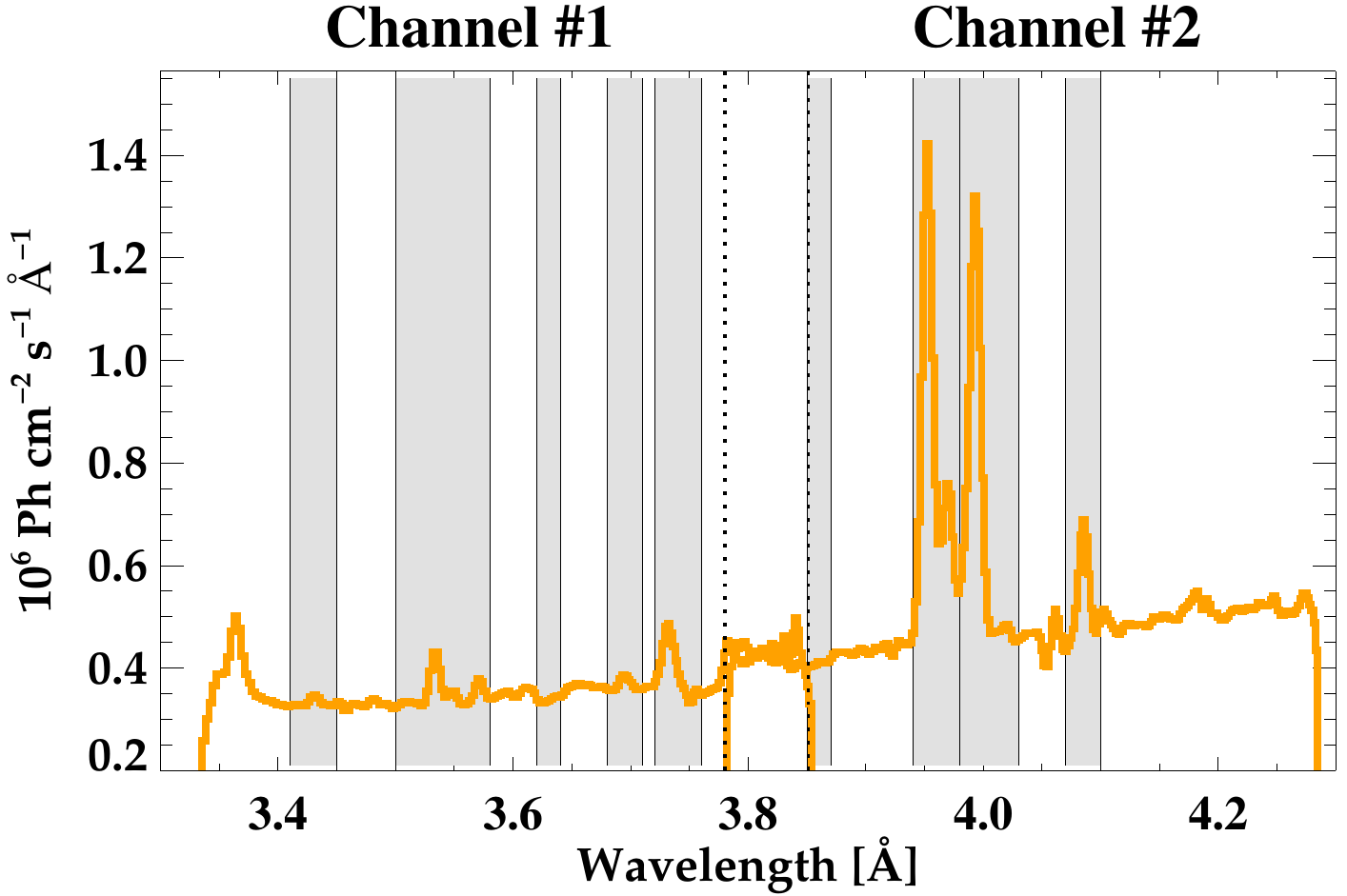}
\caption{Average spectrum observed by RESIK for the 20 September 2002 flare with selected spectral bands used for DEM analysis. The spectral range where the no. 1 and no. 2 RESIK channels  overlap is  marked by two vertical dotted lines.}
\label{FigVibStab}
\end{figure} 
\begin{table}[]
\tiny
   \centering
   \caption{Spectral bands used in calculation of DEM distributions from RESIK observations.}
\begin{tabular}{r|c|c|c}

{No.} & {Wavelength range [\AA]} & {Main line} & {Channel}\\
\hline
{1} & {3.41 -- 3.45} & {Ar {\sc{xvi}} 1s$^2$-1s3p sat.} &{1} \\
{2} & {3.50 -- 3.58} & {K {\sc{xviii}} triplet} &{1} \\
{3} & {3.62 -- 3.64} & {continuum} &{1} \\
{4} & {3.68 -- 3.71} & {S {\sc{xvi}} 5p} &{1} \\
{5} & {3.72 -- 3.76} & {Ar {\sc{xviii}} 2p} &{1} \\
{6} & {3.85 -- 3.87} & {continuum} &{2} \\
{7} & {3.94 -- 3.98} & {Ar {\sc{xvii}} triplet (w and x+y)} &{2} \\
{8} & {3.98 -- 4.03} & {Ar {\sc{xvii}} triplet (z)} &{2} \\
{9} & {4.07 -- 4.10} & {S {\sc{xv}} 4p} &{2} \\
\end{tabular}

\end{table}
For any available set of observed fluxes \textit{N} (\textit{i}=1, 2, 3, ... \textit{N})  the system of \textit{N} Fredholm's integral equations of the first kind has to be resolved. Thus the main challenge in the calculation of differential emission measure distributions is solving an inverse problem which generally lacks a unique solution \citep{Craig_1976Natur.264..340C} and has limitations due to the presence of random and systematic errors. However, using the techniques developed for this purpose and available in the literature, it is possible to determine DEM distributions and achieve stable solutions. The benchmark test given by  \cite{Aschwanden_2015SoPh..290.2733A} illustrates this point. In recent years, some new methods were introduced and described in the literature. For example, \cite{Hannah_2012A&A...539A.146H}  developed  an enhanced regularisation algorithm to be used in RHESSI X-ray spectral analysis and  applied it to observations with Hinode \citep{Golub_2007SoPh..243...63G}. \cite{Plowman_2013ApJ...771....2P} proposed a method consisting of a fast, simple regularised inversion in conjunction with an iteration scheme for the removal of residual negative emission measure. \cite{Cheung_2015ApJ...807..143C} used the concept of sparsity to develop an inversion method for DEM determination from a few data points as in the case of the Atmospheric Imaging Assembly (AIA) on board the Solar Dynamics Observatory (SDO) or X-ray telescope (XRT) onboard Hinode. This so-called sparse inversion code was modified by \cite{Su_2018ApJ...856L..17S} to use new basis functions and tolerance control. \cite{Kepa_2020SoPh..295...22K} tested and applied a  differential evolution method for determinations of DEM distributions based on RESIK data. \cite{Plowman_2020arXiv200606828P} presented and discussed a new fast and robust technique to reconstruct temperature distributions from EUV images with better performance than previous ones and with applications to future X-ray imaging spectroscopy.
\begin{figure*}[ht!]
   \includegraphics[width=15.5cm]{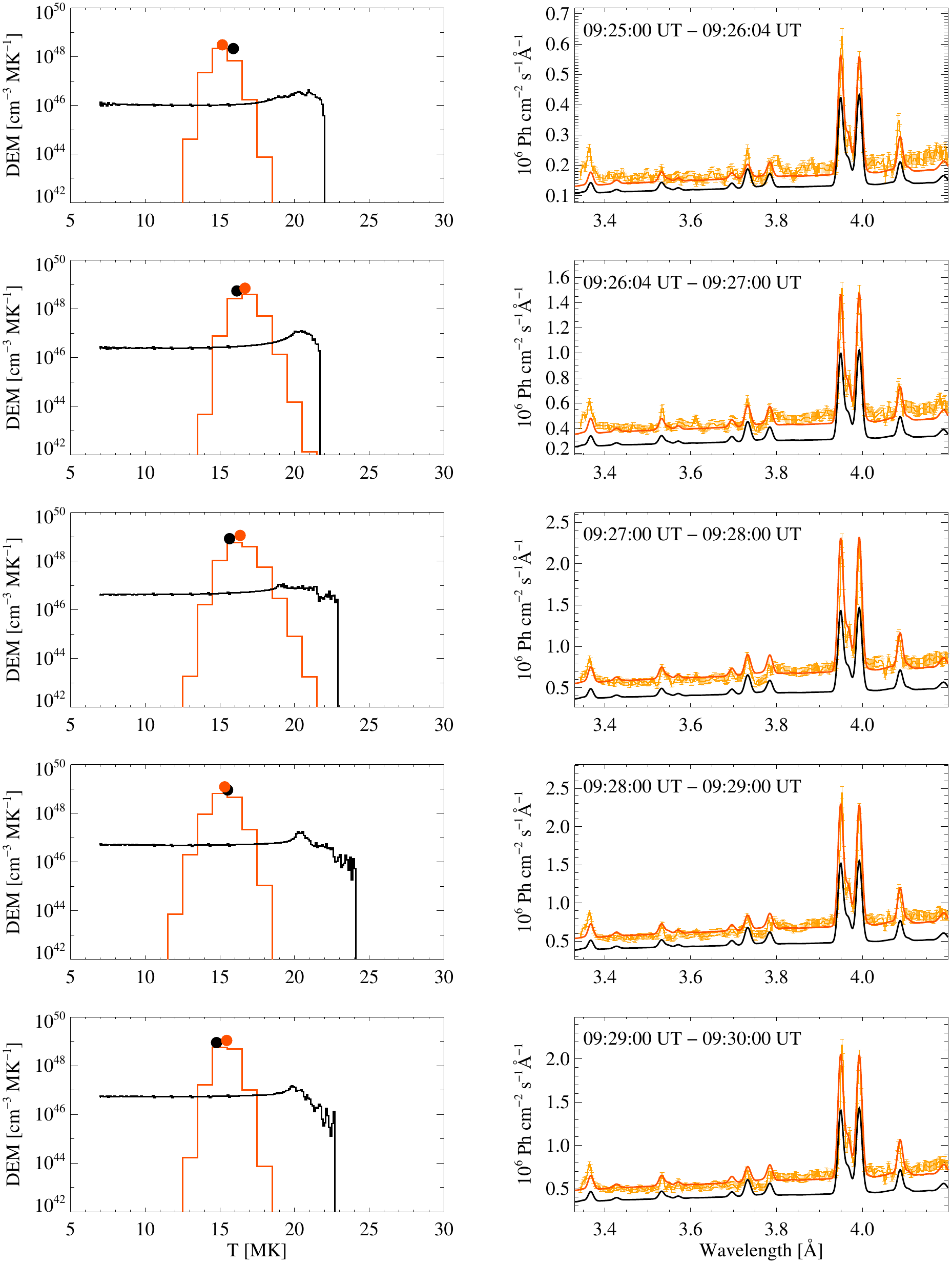}
     \centering
    \caption{Sequence of DEM distributions (left panels) and observed and calculated spectra (right) taken at five selected time intervals.  Black and orange colours are related to the 1D-HD model and RESIK DEMs, respectively. The dots correspond to the values of average temperature and total emission measure (see the text) calculated from the DEM distributions. Right panels: RESIK measured spectra with uncertainties are shown in light orange. The synthetic spectra calculated based on DEM distributions (from left panel) are plotted in black and dark orange lines.}
   \end{figure*}
Here we investigate two DEM models determined by two different approaches. Firstly, we determined the differential emission measure distributions based on RESIK spectra. We used an iterative Withbroe-Sylwester method \citep{Sylwester_B_2015ApJ...805...49S}  based on a Bayesian technique and  nine fluxes which contain  line and continuum emissions.  The average RESIK spectrum for 20 September 2002 flare and the spectral bands used in the DEM determinations are shown in Figure 4. The characteristics of selected spectral ranges (wavelength ranges and descriptions of the main lines observed in particular spectral bands) are given in Table 1. The  emission functions corresponding to the chosen spectral ranges were calculated using Chianti 8.1 \citep{Dere_1997A&AS..125..149D} package for coronal abundances \citep{1992PhyS...46..202F} and ionisation balance by \cite{Bryans_2009ApJ...691.1540B}.
        The DEM distributions obtained based on RESIK fluxes using the Withbroe-Sylwester method have been described in several previous works, such as for example  \cite{Kepa_2006SoSyR..40..294K,Kepa_2020SoPh..295...22K}, \cite{Sylwester_B_2010A&A...514A..82S, Sylwester_B_2015ApJ...805...49S}. Consequently we omit details of the method and tests and instead focus on the analysis of the 20 September flare. 

The DEM distributions can also be determined based on the 1D–HD 
model flare described in Section 3. We used the temperature, electron density, and volume dV in 3000 predefined segments (cells) of the loop to construct the DEM for each time-step of the model. The flare models were calculated every second, so we obtained DEM distributions for each second from 09:18:15 UT to 09:41:28 UT. Next, we calculated average DEMs distributions for the same time intervals as those for which the DEMs from RESIK data were determined. This allowed us to compare the DEM models determined based on RESIK and RHESSI data using two different approaches.

The results obtained for five selected time intervals (during the flare) are shown in Figure~5 (left panel). Differential emission measures resulting from the 1D-HD modelling are shown in black, with RESIK data in orange. It can be seen that the DEM distributions determined based on RESIK spectra are one-component distributions with plasma temperature in the range of 14~MK~--~19~MK. Such distributions are related with using fluxes in spectral bands containing high-temperature argon, sulphur, and potassium lines. The DEM distributions from 1D-HD models were constructed for temperatures above 6 MK. For temperatures lower then 6~MK the contribution for RESIK flux (for spectral range 3.33~\AA~--~4.32~\AA~or 3.00~keV~--~3.66~keV) was found to be negligible (about 10$^{-6}$).

For each of the selected time intervals, the values of average temperature ($T_a$) and total emission measure ($EM_{tot}$) (black and orange dots in the left panel of Figure 5 and orange dots in Figure 6) were calculated. These were determined based on the DEM distributions (both these related to RESIK data and those obtained from 1D-HD modelling results) using the following formula: $EM_{tot}=\int_{T}\varphi(T){dT}$ and $T_{a}=\frac{\int_{T}T\varphi(T){dT}}{\int_{T}\varphi(T){dT}}$.

\begin{figure}
\centering
\includegraphics[width=9cm]{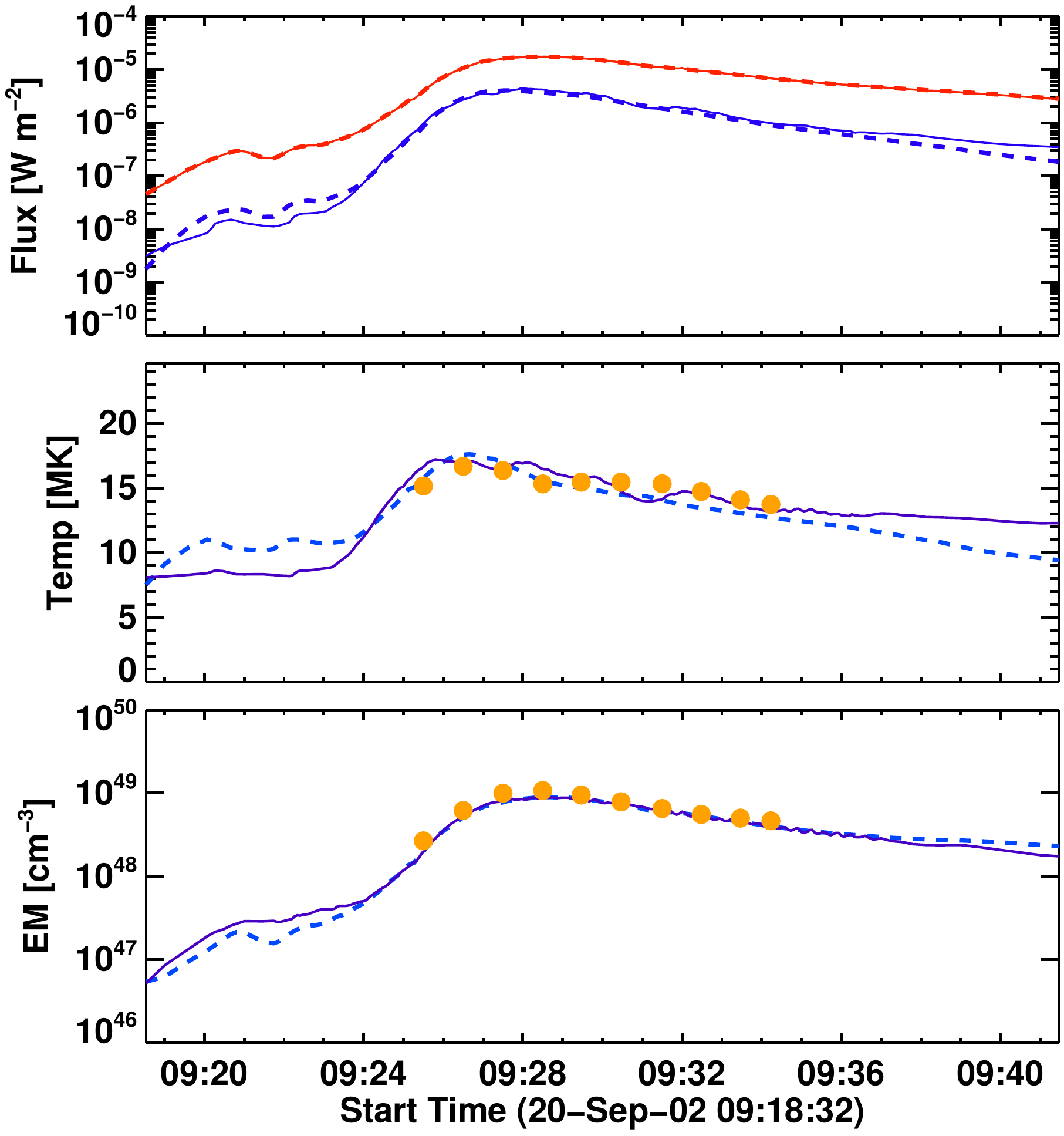}
\caption{GOES light curves (top), temporal evolution of temperature (middle panel),  and emission measure  (bottom) calculated based on GOES fluxes. The values related with  observations from GOES are plotted as dashed lines, and solid lines represent the values as obtained based on GOES fluxes synthesised using the numerical 1D-HD model. Red and blue colours in the top panel correspond to the GOES 0.5~--~4~\AA~ and 1~--~8~\AA~ channels. The orange dots  represent the average temperature ($T_a$, middle panel) and  total emission measure ($EM_{tot}$, bottom) values calculated based on the DEM distributions from RESIK data.
}
\label{FigVibStab}
\end{figure} 
\begin{table*}
 \begin{center}
 \caption{List of analysed flares.}
 \vspace{-0.2cm}
\begin{tabular}{c |c c c c c} 
\hline\hline
No. & Date & Maximum [UT] & GOES class & Time of observations [UT] & colour in Figure 7\\
\hline
1 & {15 Apr. 2002} & 03:55 & M1.2 & 03:06:10 -- 03:37:48 & blue\\
2 & {16 Aug. 2002} & 22:12 & M1.2 & 22:10:30 -- 22:12:06  &  orange\\
{3} & {29 Sep. 2002} & {06:39} & {M2.6}  & {06:35:38 -- 06:36:50} & pink\\
{4} & {14 Nov. 2002} & {22:26} & {M1.0} & {22:25:06 -- 22:26:25} & green\\
{6} & {3 Jan. 2003} & {20:50} & {C2.0}  & {20:49:38 -- 20:59:59} & yellow\\
{7} & {7 Jan. 2003}  & {04:35} & {C1.6}  & {04:27:46 -- 04:33:41} & purple\\
{8} & {7 Jan. 2003} & {11:12} & {C2.9}  & {10:45:22 -- 10:58:30} & brown\\
\end{tabular}
\end{center}
    \label{intSt}
\end{table*}
 It can be seen that the average temperatures  and total  emission measure values  obtained based on two different DEM distributions are not exactly the same. Total emission measures obtained based on the results of hydrodynamic modelling  are systematically lower than those calculated from RESIK fluxes. This relationship is even more apparent in the right panel of Figure 5, where the comparison of average spectra  observed by RESIK (in light orange) and synthetic spectra determined based on DEMs from numerical models (in black) and based on RESIK DEM distributions (dark orange) are shown. The synthetic spectra corresponding to 1D-HD modelling for all cases  do not accurately describe the observed RESIK spectra. This discrepancy is probably related to errors in cross-calibration of RESIK and RHESSI instruments and is discussed in the Section 5. 

The main results of modelling for the 20 September 2002 flare are given in Figure 6. The synthesised GOES light curves in 0.5~--~4~\AA~ and 1~--~8~\AA~ bands, which directly result from our modelling,   closely  follow those observed for the analysed flare. The model parameters were determined for the achievement of the best agreements between synthesised and observed GOES fluxes in the 1~--~8~\AA~ band. Therefore, the real test for the accuracy of the model is the comparison of the 0.5~--~4~\AA~ band. Smaller differences between synthesised and observed light curves in the 0.5~--~4~\AA~ band correspond to a better representation of the physical conditions of the flaring plasma by the 1D-HD model/simulation. At the beginning of the flare, the synthesised light curve (solid blue colour) is underestimated  (blue dashed line); from 09:24:00~UT up to 09:36:30~UT it has a similar pattern. During the decay phase (after 09:36:30~UT), the synthesized curve is located above the observed one until  to the end of the simulation. The calculated light curves for the analysis event did not differ significantly from the observed ones especially for the time interval in which RESIK observations were available. It seems that our model correctly simulates the main physical processes despite the large simplifications that are used. 

A comparison was also made between the average temperature and total emission measure values determined from RESIK DEM distributions with the temperatures and emission measure values as calculated using synthesised and observed GOES fluxes. The results are shown in the middle and bottom panels of Figure 6. The temporal behaviours of $T_a$ and $EM_{tot}$ values obtained based on RESIK DEM distributions are very similar to those  from GOES observations and hydrodynamic models. The mean temperatures values calculated using RESIK (based on DEM distribution) and GOES (isothermal assumption) data agree with each other to within 0.5~--~2~MK.  The values of the total emission measure obtained based on RESIK DEM distributions are systematically higher than
those from GOES. 
        
The difference between plasma parameters obtained based on the RESIK observations and from the 1D-HD flare model may be the result of several factors. The first is that the modelling of the flare was based on RHESSI observations and this instrument has a different temperature sensitivity from that of  RESIK. The second reason is related to the fact significant uncertainties are associated with RHESSI measurements below 4~keV   \citep{Smith_2002SoPh..210...33S}.

\section{RHESSI and RESIK correlation}

The lower end of the RHESSI spectral energy range (down to 3 keV) overlaps with RESIK channels no.~1 and no.~2 offering a highly useful means of cross-calibration. However, one must note that the RHESSI effective area abruptly decreases below 6 keV. RHESSI does not record emission of the thermal plasma below 3 keV, and the quality of the recorded data is low between 3 keV and 6 keV. This property is important when comparing fluxes from RHESSI and RESIK instruments.

\cite{Chifor_2007A&A...462..323C} compared observations from GOES, RESIK, and RHESSI for one flare observed on 22 February 2003. These latter authors obtained good agreement in temperatures achieved during the later decay phase of the flare. \cite{Dennis_2004cosp...35.1284D} studied  the flare observed on 26 April 2003 in more detail and reported  agreement between the  RESIK continuum and the RHESSI spectrum at around 3.5~keV to within about 20\%. Correlations between RESIK and RHESSI data for individual  events without a detailed analysis is also presented in several other  papers, such as for example \cite{Sylwester_2005SoSyR..39..479S} and \cite{Gburek_2008AdSpR..42..822G}. 

In our study we selected a group of seven C and M GOES class  flares which were simultaneously observed with RESIK (level 2 data) and RHESSI (RHESSI in A0 attenuator state) and occurred between 2002 and 2003. The main characteristics of these events, such as times of maximum, GOES classes, and time periods in which observations have been compared are given in Table~2.

We analysed the spectra of these events. We reconstructed the RHESSI photon spectra in the OSPEX environments by fitting the count data to the bremsstrahlung  spectrum. The methodology was similar to the 20 September 2002 flare analysis. We assumed an isothermal approximation of plasma and power-law energy distribution. The background emission determined using a linear fit to the measurements before and/or after the flare was subtracted from the flare data. The RHESSI spectra have been reconstructed in the same way for each time intervals for which RESIK measurements were available. For each interval corresponding to an individual RESIK spectrum, the total flux observed by RESIK and RHESSI in the energy range 3.00~keV~--~3.66~keV (3.38~\AA~--~4.13~\AA) was calculated.
\begin{figure}[h!]
\centering
\includegraphics[width=9cm]{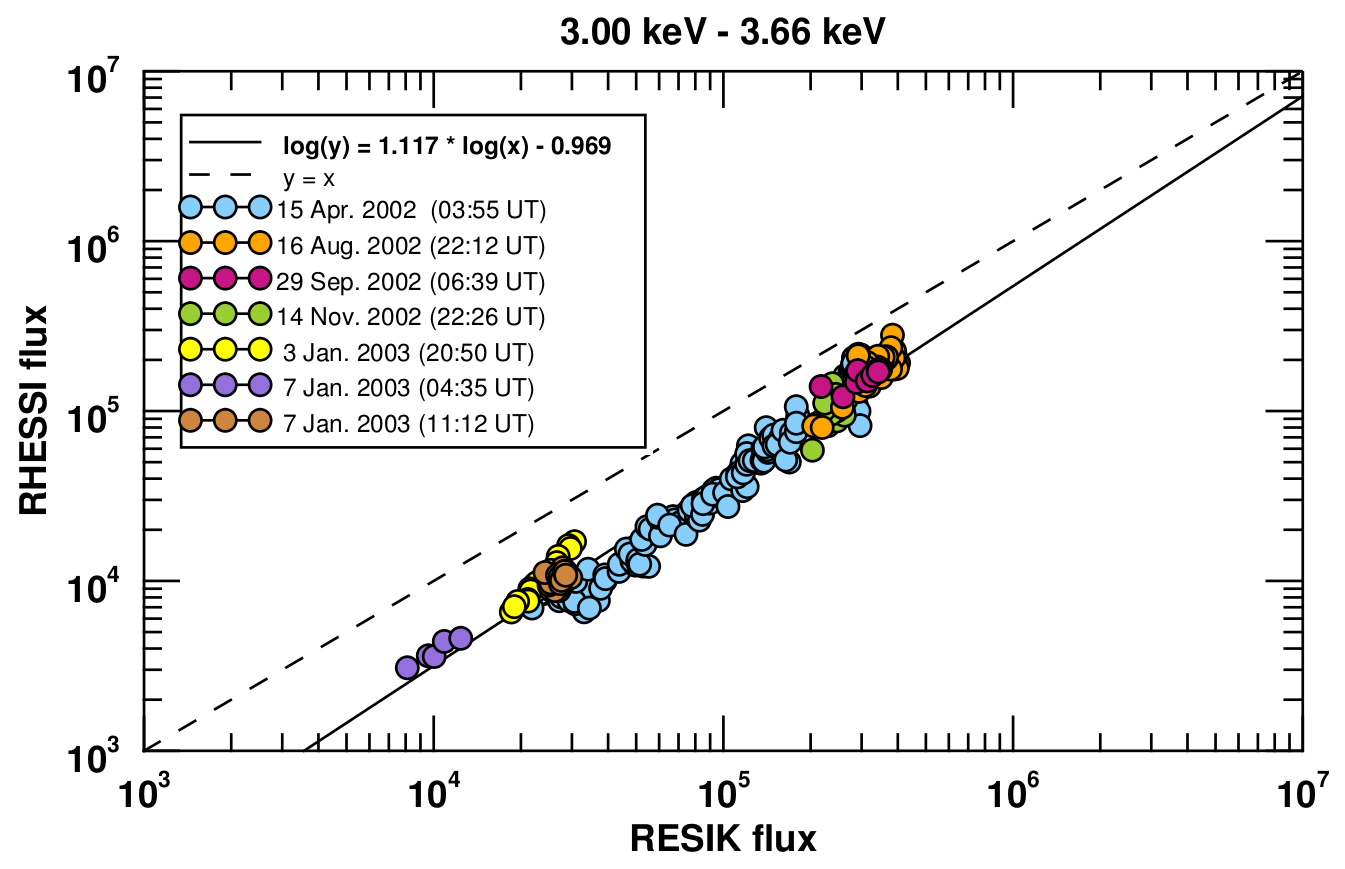}
\caption{Correlation of RESIK and RHESSI fluxes for energy range 3.00~--~3.66~keV. Different colours represent observations from different flares (see Table~2). The analysis of RHESSI data was made applying grids 1F, 3F, 4F, 6F, 8F, and 9F. The continuous black line represents the linear regression, the coefficients of which are calculated by the least squares method. The dashed black line represent equal fluxes.}
\label{FigVibStab}
\end{figure}
The correlation of RHESSI and RESIK fluxes in the energy range 3.00~--~3.66~keV obtained based on nearly 350 spectra of flares from Table~2 is shown in Figure~7. The different colours correspond to different events. The dashed black line represents equal values. 

The apparent scatter of points is probably related to the absolute calibration of the Detector Response Matrix matrix (DRM) of RHESSI instrument in low energies, which directly contributes to differences in derived spectral values. It could be associated also with determination and subtraction of background level, which is a key issue of RHESSI data analysis. Despite the  above problems, RHESSI and RESIK fluxes are related by a straight-line when plotted logarithmically (Figure~7). This line is not parallel to the straight dashed line corresponding to equal RHESSI and RESIK fluxes). Also, the correlation is weaker for small fluxes.

  
\begin{figure*}
\hspace{.5cm}
 \begin{minipage}[t]{0.5\linewidth}   
    \includegraphics[width=8cm]{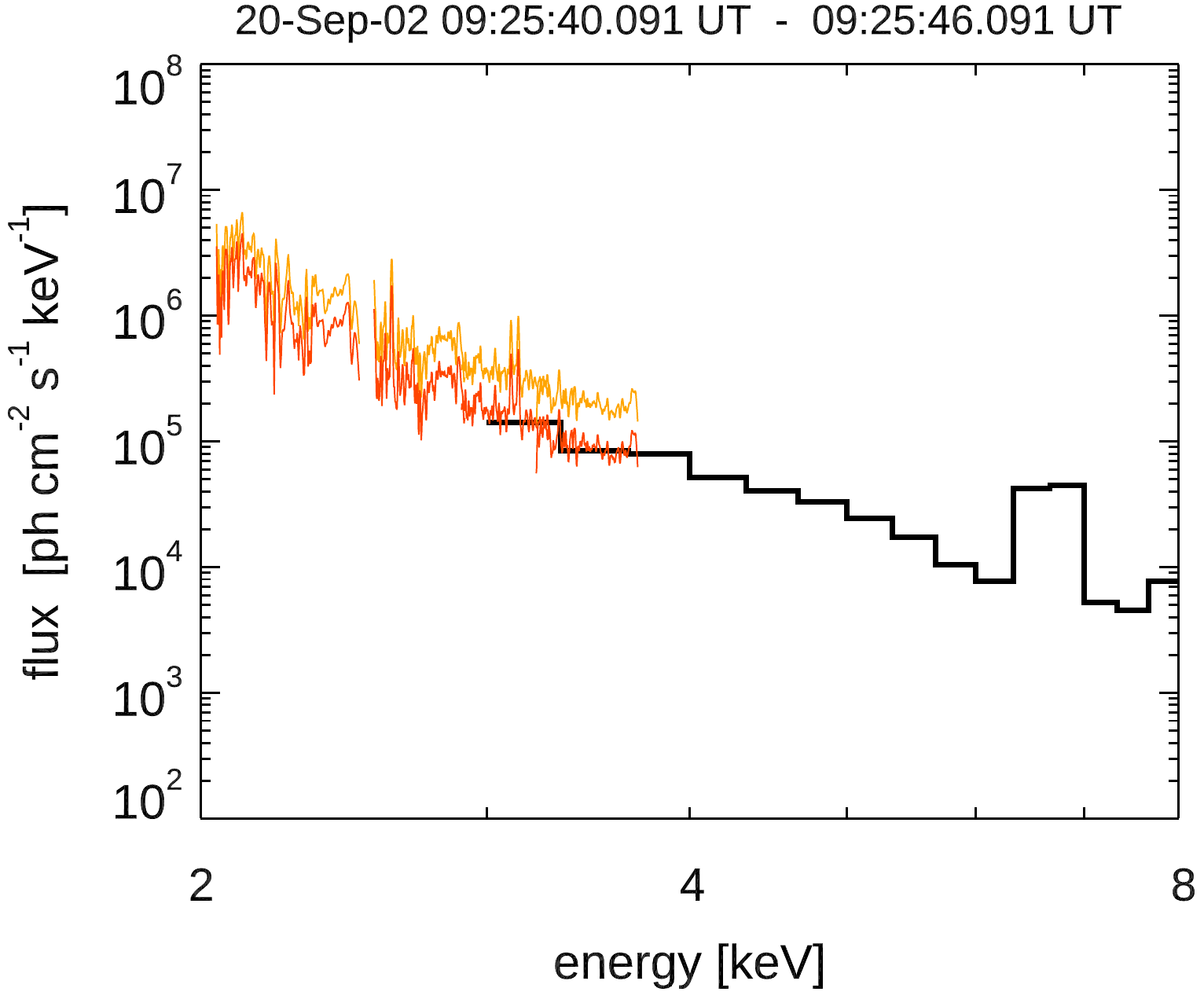}
\end{minipage} 
\vspace{0.3cm}  
\hspace{-.5cm}
\begin{minipage}[t]{0.5\linewidth}
   \includegraphics[width=8cm]{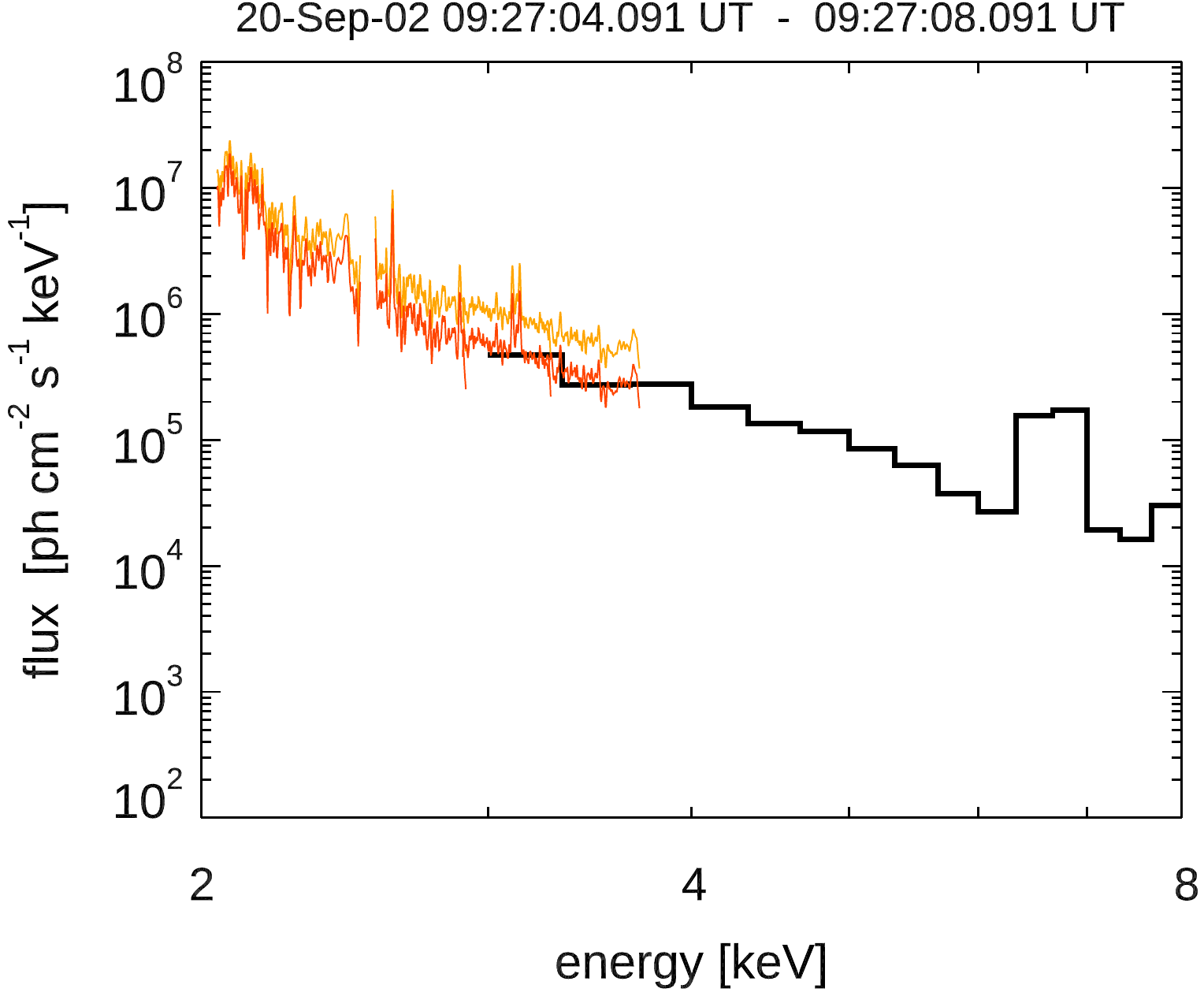}
\end{minipage}
\vspace{0.3cm} 
\hspace{.5cm} 
\begin{minipage}[t]{0.5\linewidth}   
    \includegraphics[width=8cm]{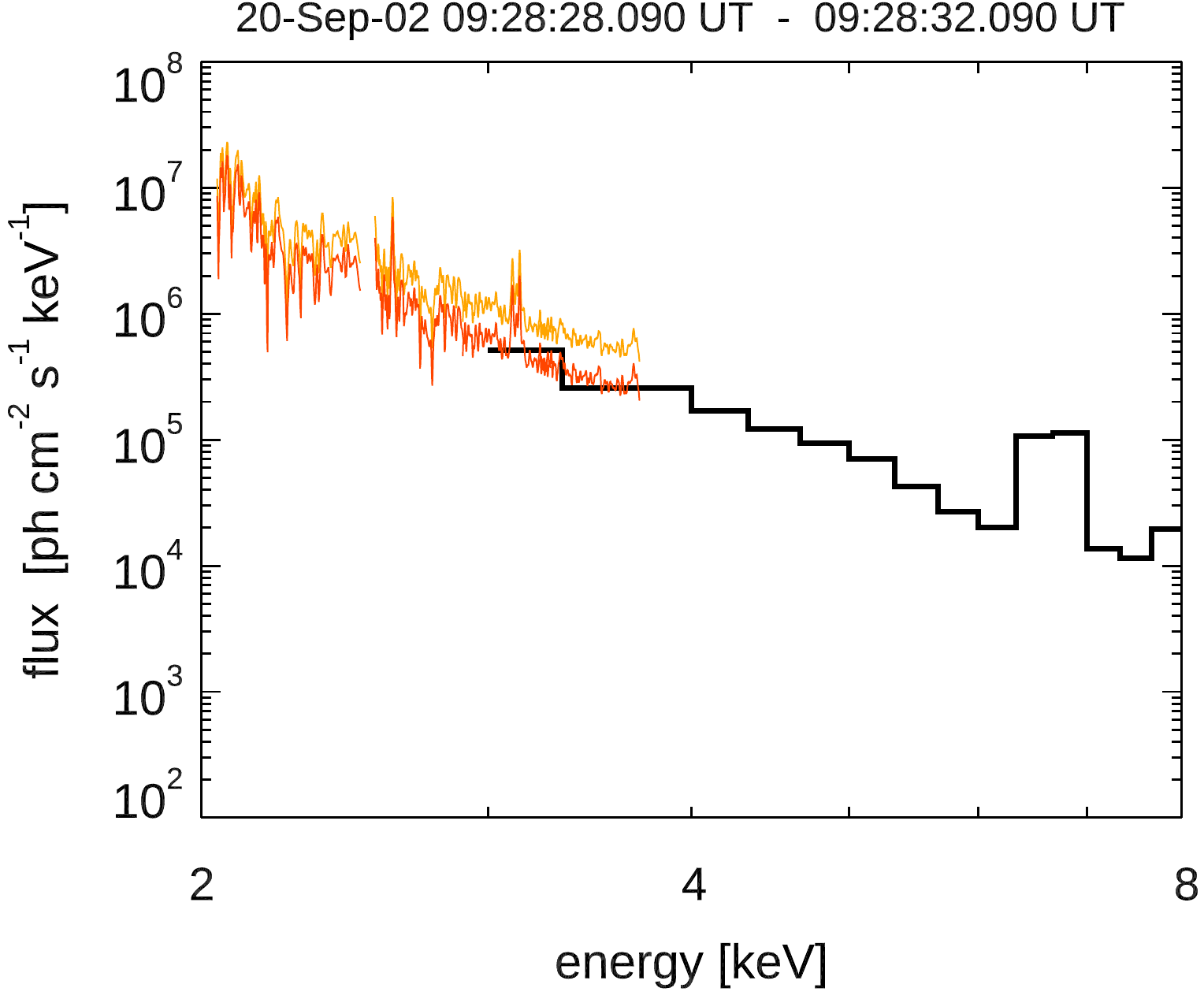}
\end{minipage}
\hspace{-.5cm} 
 \begin{minipage}[t]{0.5\linewidth}
   \includegraphics[width=8cm]{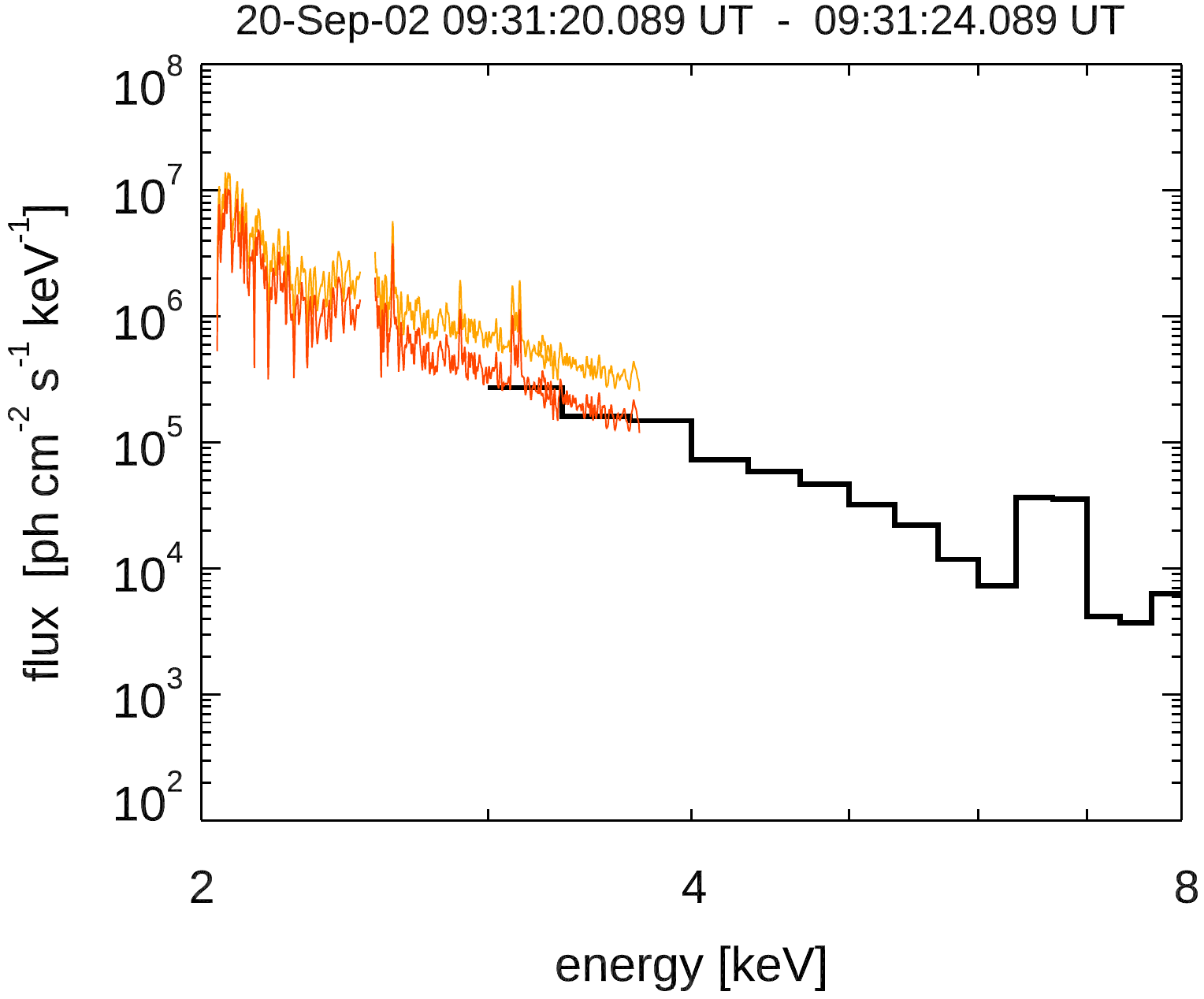}
\end{minipage}
      \caption{Example comparison of the four RHESSI and RESIK observed spectra.  `Original' and `multiplied' RESIK spectra are shown in light orange and dark orange respectively. The RHESSI spectra are plotted in black.}
         \label{FigVibStab}
   \end{figure*}    
\begin{figure*}
\centering
\includegraphics[width=16cm]{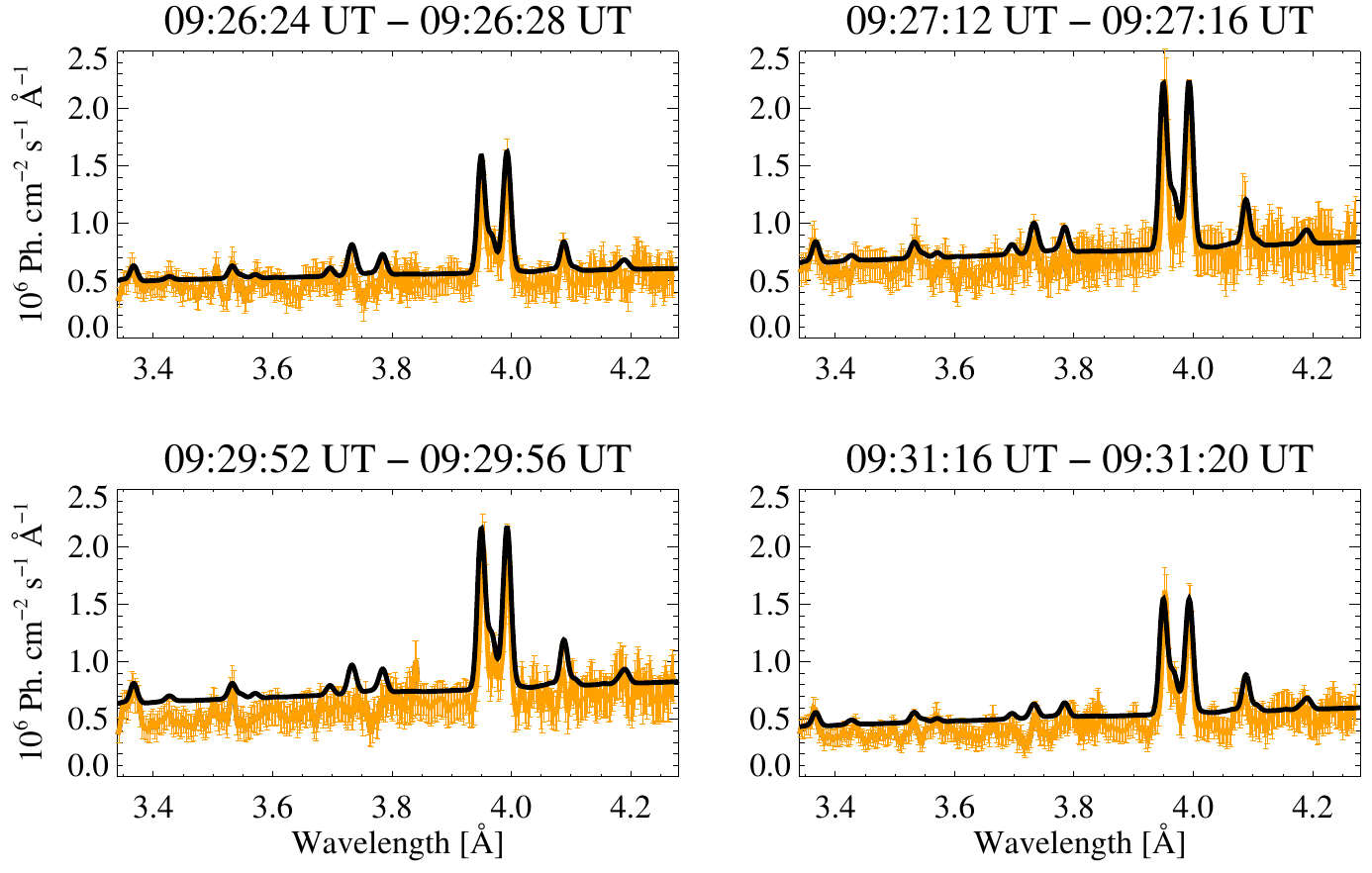}
\caption{Comparison of RESIK observations for 20 September 2002 flare (in orange) and synthetic spectra (in black) calculated based on the results of 1D-HD modelling and multiplied by the correlation coefficients for four selected  observations taken in the different phases of the flare's evolution. }
\label{FigVibStab}
\end{figure*}     

The relationship shown in Figure 7 can be used to explain the difference between RHESSI 
and RESIK observations for the flare of 20 September 2002. For this purpose, we multiplied the `original' RESIK spectra according to the relation from Figure 7. In Figure 8 we present an example of such a comparison for the four spectra observed in different phases of the flare. Original and `multiplied' RESIK spectra are plotted in light and dark orange, respectively. The RHESSI spectra are shown in black. After taking into account the correlations between RESIK and RHESSI observations, the spectra from both instruments overlap in the energy region from 3.00 to 3.66~keV.

Finally, the relationship between RHESSI and RESIK fluxes were used to the explain differences between observed and calculated (based on DEM distribution from the 1D-HD model) spectra. Figure~9 shows four selected individual  RESIK observations of the flare on 20 September 2002 (in orange) and corresponding  synthetic spectra (in black) calculated based on the results of the modelling while  the correlation coefficients were applied. The synthetic spectra (in black) correctly reproduce the observed ones (within observational errors), although they were calculated based on the simple one-loop model and many other assumptions, which have been made during 1D-HD modelling of the flare.

\section{Conclusions}

One of the fundamental assumptions when modelling a solar flare is related to the geometry of flaring loops. Solar flaring loops are 3D objects with internal structures \citep{Aschwanden_2001SoPh..204...91A,Aschwanden_2006cosp...36.3417A}. Unfortunately, even  modern ground-based and satellite-based instruments are not able to resolve their structures exactly. These observational limitations cause fundamental problems in accurate numerical modelling of the flares. Thus, there is no means as yet to investigate possible subsecond heating episodes or the thermodynamic evolution of substructures that are beyond the spatial resolution of RHESSI. These limitations require simplified models of flares.  Simplifications of the real geometry of the flare influence the accuracy of calculated numerical modelling. However, discrepancies between real flaring structures and models of flaring loops may also be related to models of the energy deposition mechanism inside the threads and the assumed mechanisms of transfer and redistribution of the energy already deposited in the flaring loops. The physics behind processes involved in abrupt heating of flaring plasma is still not fully understood and needs to be revised for example by a comparison of observational features of the flaring transient plasma with the results of numerical modelling of the flares. The 1D-HD models that are relatively simple and require a moderate volume of the necessary calculations appear to be an extremely valuable tool for studying the physics of flares. 

Another assumption in flaring loop models is that of a thick-target model in which a beam of electrons is injected at the top of a loop and   `precipitates' downwards in the solar atmosphere. This model appears to be accurate for most observations and is usually used for the analysis of HXRs. In the case of the flare studied here, the thick-target model explains the observational facts very well. However, it should be noted  that there are  other scenarios explaining  the  acceleration of electrons and/or chromospheric evaporation in the literature. For example, \cite{Battaglia_2009A&A...498..891B}  developed an alternative mechanism to electron-beam-driven evaporation, namely conductively  driven evaporation. These latter authors performed a detailed study of the pre-flare phase of four solar flares using imaging and spectroscopy from the RHESSI satellite and explained the time evolution of the observed emission for all analysed events as an effect of saturated heat flux. It is also worth mentioning  the idea of the `collapsing trap' \citep{Somov_1997ApJ...485..859S,Jakimiec_2002ESASP.506..645J} which is proposed as an efficient accelerator of particles. A collapsing trap occurs as a result of moving (collapsing) magnetic field lines and plasma trapped between magnetic field `mirrors'. An alternative mechanism for particle acceleration is also plasma turbulence, which has been  studied by several authors, such as for example \cite{Miller1987SoPh..113..195M}, \cite{Hamilton_1992ApJ...398..350H}, and \cite{Lenters_1998ApJ...493..451L}. Recently, \cite{Hudson_2020} presented evidence of a hot X-ray `onset'  interval of enhanced isothermal plasma temperatures in the range of 10 -- 15~MK up to tens of seconds prior to the flares impulsive phase. This `hot onset' interval occurs during the initial SXR increase and prior to the detectable HXR emission. These HXR onsets appear before there is evidence of collisional heating by non-thermal electrons, and therefore they challenge the standard flare heating modelling techniques.

In this article we reconstructed the RESIK SXR spectra based on the results of a 1D-HD model of a flaring loop. The modelling and analysis were performed for the M1.8 GOES class flare on 20 September 2002 . Estimations of the geometrical parameters of the flaring loop were made based on the RHESSI and SOHO/EIT images. Because this flare had a very simple, one-loop structure, we were able to use the hydrodynamic 1D-HD model  for the flaring plasma study. The distribution of non-thermal electrons determined from RHESSI HXR spectra was used as input data for the model calculations. According to the flare model, these non-thermal electrons travel down from the loop top and give rise to heating and chromospheric evaporation.  The evaporated plasma radiates in SXRs. The total energy of electrons was controlled by comparison of the observed and calculated fluxes in the 1~--~8~\AA~channel from GOES data. 

We determined and  compared  the physical properties of the emitting plasma using two independent techniques  for DEM calculations: The Withbroe-Sylwester likelihood approach for analysis of RESIK spectra and the temperature--density distributions obtained as the result of flare modelling. The DEM distributions obtained using these two approaches differ in shape, although the average temperatures and total emission measure values maintained based on these distributions are similar.  

We also calculated thermal spectra which should be observed from this loop based on the 1D-HD  model and compared them with RESIK observations in 3.33~\AA~- 4.32~\AA.  The reconstructed RESIK spectra  calculated in this manner for the 20 September 2002 flare are consistent within a factor of approximately two with observations throughout  most of the duration of the flare. Similar results and conclusions were obtained when  data from Yohkoh/HXT and Yohkoh/BCS were compared \citep{Falewicz_2009A&A...508..971F}. These latter authors determined the time series of the spectra of three solar flares at various moments of their evolution (from the beginning of the impulsive phases to beyond maxima of the X-ray emission) using a 1D-HD numerical model of the solar flare and standard software to calculate BCS synthetic spectra of the flaring plasma. The models of the flares were calculated using observed energy distributions of the non-thermal electron beams injected into the loops. The synthesised BCS spectra of the flares were compared with the relevant observed BCS spectra. This allowed many observational aspects to be explained, such as for example the stationary component of the spectrum which should be observed for almost all flares during their early phases of evolution. By contrast, the blueshifted component resulting from the motion of the plasma along the loop may be completely invisible in terms of its geometric effect, for example when the motion along the loop (even with high velocity) is perpendicular to the line of sight.

Finally, to explain the difference between observations and the synthetic spectra calculated from the 1D-HD model, we determined the cross-correlation coefficients between the RESIK
and RHESSI instruments. Their application allowed good agreement to be achieved between RESIK
and RHESSI observations. The  RHESSI, RESIK, and GOES instruments are
well cross-calibrated, giving similar parameters for observed phenomena.

The results obtained during this study  allow us to conclude that our model of flaring-loop heating with NTE beams  has strong observational and theoretical foundations. Despite the simplifications used during the modelling, the results of our simulations are consistent with observations. The one-loop approximation works in many flares where images (mainly in SXRs or EUV) suggest a single-loop configuration. This means that the basic physics involved is understood and is accurately represented by the model, especially processes of transport of NTE electrons down through the loop, their dissipation in the chromosphere (Fisher or Focker-Plank approximation), and processes of chromospheric evaporation.

\begin{acknowledgements}
 We would like to thank anonymous reviewer for his insightful comments and valuable suggestions. We also thank Barbara Sylwester, Janusz Sylwester  and Kenneth J.H. Phillips for corrections and helpful comments. We acknowledge financial support from the Polish National Science Centre grants No. 2017/25/B/ST9/01821 and 2015/19/ST9/02826.
\end{acknowledgements}

\bibliography{mybibfile}
\end{document}